\documentclass [12pt]{article}
\usepackage[round]{natbib}

\usepackage {graphicx}
\usepackage{epsfig,amsmath,latexsym,amssymb}
\usepackage{bm}
\usepackage{caption,placeins}
\usepackage{booktabs}
\usepackage{hyperref}
\usepackage{multirow,color}

\usepackage[margin=0.9in]{geometry}
\newtheorem{remark}{Remark}[section]
\newtheorem{proposition}{Proposition}[section]

\begin{document}

\title{\vspace{-1.5cm} \textbf{Should the advanced measurement approach be replaced with the standardized measurement approach for operational risk?}}
\author{\textbf{Gareth W. Peters}\\
{\emph{Department of Statistical Sciences, University College London, UK}} \\\small{\emph{email: gareth.peters@ucl.ac.uk}}\\
\textbf{Pavel V.~Shevchenko}\footnote{corresponding author} \\
\emph{CSIRO Australia}\\ \small{\emph{e-mail: Pavel.Shevchenko@csiro.au}}\\
\textbf{Bertrand Hassani}\\
\emph{Universit\'{e} Paris 1 Panth\'{e}on-Sorbonne}\\ \small{\emph{email: bertrand.hassani@univ-paris1.fr}}\\
\textbf{Ariane Chapelle}\\
\emph{Department of Computer Science, University College London UK}\\ \small{\emph{email: a.chapelle@ucl.ac.uk}}}

\date{{Draft paper: 1 July 2016}}

\maketitle

%\title{Should AMA be Replaced with SMA for Operational Risk?}
%
%\author[label1]{Gareth W. Peters\corref{mycorrespondingauthor}}
%\ead{gareth.peters@ucl.ac.uk}
%\author[label2,label1]{Pavel V.~Shevchenko}
%\ead{Pavel.Shevchenko@csiro.au}
%\author[label3]{Bertrand Hassani}
%\ead{bertrand.hassani@univ-paris1.fr}
%\author[label4]{Ariane Chapelle}
%\ead{a.chapelle@ucl.ac.uk}
%
%\address[label1]{Department of Statistical Sciences, University College London, UK}
%\address[label2]{CSIRO Australia}
%\address[label3]{Universit\'{e} Paris 1 Panth\'{e}on-Sorbonne}
%\address[label4]{Department of Computer Science, University College London UK}

\begin{abstract}
\noindent Recently, Basel Committee for Banking Supervision proposed to replace all approaches, including Advanced Measurement Approach (AMA), for operational risk capital with a simple formula referred to as the Standardised Measurement Approach (SMA). This paper discusses and studies the weaknesses and pitfalls of SMA such as instability, risk insensitivity, super-additivity and the implicit relationship between SMA capital model and systemic risk in the banking sector. We also discuss the issues with closely related operational risk Capital-at-Risk (OpCar) Basel Committee proposed model which is the precursor to the SMA. In conclusion, we advocate to maintain the AMA internal model framework and suggest as an alternative a number of standardization recommendations that could be considered to unify internal modelling of operational risk. The findings and views presented in this paper have been discussed with and supported by many OpRisk practitioners and academics  in Australia, Europe, UK and USA, and recently at OpRisk Europe 2016 conference in London.\\

\vspace{0.2cm}
\noindent \textbf{Keywords:} \emph{operational risk, standardised measurement approach, loss distribution approach, advanced measurement approach, Basel Committee for Banking Supervision regulations.}\\
\\
\noindent\emph{JEL classification:} G28, G21, C18
\end{abstract}

\newpage

%%%%%%%%%%%%%%%%%%%%%%%%%%%%%%%%%%%%%%%%%%%%%%%%%%%%%%%%%%%%%%%%%%%%
%%%%%%%%%%%%%%%%%%%%%%%%%%%%%%%%%%%%%%%%%%%%%%%%%%%%%%%%%%%%%%%%%%%%
\section{Introduction} \label{sec:introduction}
%%%%%%%%%%%%%%%%%%%%%%%%%%%%%%%%%%%%%%%%%%%%%%%%%%%%%%%%%%%%%%%%%%%%
%%%%%%%%%%%%%%%%%%%%%%%%%%%%%%%%%%%%%%%%%%%%%%%%%%%%%%%%%%%%%%%%%%%%
\noindent Operational risk (OpRisk) management is the youngest of the three major risk branches, the others being market and credit risks within financial institutions. The term OpRisk became more popular after the bankruptcy of the Barings bank in 1995, when a rogue trader caused the collapse of a venerable institution by placing bets in the Asian markets and keeping these contracts out of sight of management. At the time, these losses could be classified neither as market nor as credit risks and the term OpRisk started to be used in the industry to define situations where such losses could arise. It took quite some time until this definition was abandoned and a proper definition was established for OpRisk. In these early days, OpRisk had a negative definition as ``\emph{any risk that is not market or credit risk}'', which was not very helpful to assess and manage OpRisk. Looking back at the history of risk management research, we observe that early academics found the same issue of classifying risk in general, as \cite{crockford1982bibliography} noticed:
``\textsl{Research into risk management immediately encounters some basic problems of definition. There is still no general agreement on where the boundaries of the subject lie, and a satisfactory definition of risk management is notoriously difficult to formulate}''.

One thing that is for certain is that as risk management started to grow as a discipline, regulation also began to get more complex to catch up with new tools and techniques. It is not a stretch to say that financial institutions have always been regulated one way or another given the risk they bring to the financial system. Regulation was mostly on a country-by-country basis and very uneven, allowing arbitrages. As financial institutions became more globalized, the need for more symmetric regulation that could level the way institutions would be supervised and regulated mechanically increased worldwide.

As a consequence of such regulations, there has been in some areas of risk management, such as market risk and credit risk, a gradual convergence or standardization of best practice which has been widely adopted by banks and financial institutions. In the area of OpRisk modelling and management, such convergence of best practice is still occurring due to multiple factors such as: \emph{many different types of risk process modelled within the OpRisk framework; different influences and loss experiences in the OpRisk categories in different banking jurisdictions; and the very nature of OpRisk being a relatively immature risk category compared to market and credit risk.}

The question therefore arises, how can one begin to induce a standardization of OpRisk modelling and capital calculation under Pillar I of the current banking regulation accords from the Basel Committee for Banking Supervision (BCBS). It is stated under these accords, that the basic objective of the Basel Committee work has been to close gaps in international supervisory coverage in pursuit of two basic principles: that no foreign banking establishment should escape supervision; and that supervision should be adequate. It is this second note that forms the context for the new proposed revisions to the simplify OpRisk modelling approaches brought out in a sequence of two consultative documents:
\begin{itemize}
\item most recently, the Standardised Measurement Approach (\textbf{SMA}) proposed in the Basel Committee consultative document ``Standardised Measurement Approach for operational risk'' (issued in March 2016 for comments by 3 June 2016), \cite{BCBSd3552016}; and
\item closely related OpRisk Capital-at-Risk (\textbf{OpCar}) model proposed in the Basel Committee consultative document ``Operational risk - Revisions to the simpler approaches'' issued in October 2014, \cite{BCBSd2912014}.
\end{itemize}
In the consultative document \cite[page 1]{BCBSd2912014} it is noted that
``\textsl{Despite an increase in the number and severity of operational risk events during and after the financial crisis, capital requirements for operational risk have remained stable or even fallen for the standardised approaches}''. Consequently, as a result it is reasonable to reconsider these measures of capital adequacy and to consider if they need further revision. This is exactly the process undertaken by the Basel Committee in preparing the revised proposed frameworks that are discussed in this manuscript. Before, getting to the revised framework of the SMA capital calculation it is useful to recall the current best practice in Basel regulations.

Many models have been suggested for modelling OpRisk under the Basel II regulatory framework,  \cite{BaselII2006}. Fundamentally, there are two different approaches considered:
\emph{\textbf{the top-down approach}} and \emph{\textbf{the bottom-up approach}}.
A top-down approach quantifies OpRisk without attempting to identify the events or causes of losses explicitly. It can include the Risk Indicator models that rely on a number of OpRisk exposure indicators to track OpRisks and the Scenario Analysis and Stress Testing models that are estimated based on the what-if scenarios. Whereas a bottom-up approach quantifies OpRisk on a micro-level being based on identified internal events. It can include actuarial type models (referred to as the Loss Distribution Approach) that model frequency and severity of OpRisk losses.

Under the current regulatory framework for OpRisk \cite{BaselII2006}, banks can use several methods to calculate OpRisk capital: \emph{Basic Indicator Approach} (\textbf{BIA}), \emph{Standardised Approach} (\textbf{TSA}) and \emph{Advanced Measurement Approach} (\textbf{AMA}). Detailed discussion of these approaches can be found in \cite[chapter 1]{CruzPetersShevchenko2015}. In brief, under the BIA and TSA the capital is calculated as simple functions of gross income ($GI$):
\begin{eqnarray}
K_{BIA}&=&\alpha\frac{1}{n}\sum_{j=1}^3 \max\{GI(j),0\},\quad n= \sum_{j=1}^3 1_{\{GI(j)>0\}},\quad \alpha=0.15;\\
K_{TSA}&=&\frac{1}{3}\sum_{j=1}^3 \max\left\{\sum_{i=1}^8 \beta_i GI_i(j), 0\right\},
\end{eqnarray}
where $1_{\{\cdot\}}$ is the standard indicator symbol equals one if condition in ${\{\cdot\}}$ is true and zero otherwise. Here, $GI(j)$ is the annual gross income of a bank in year $j$, $GI_i(j)$ is the gross income of business line $i$ in year $j$, and $\beta_i$ are coefficients in the range $[0.12-0.18]$ specified by Basel Committee for 8 business lines. These approaches have very coarse level of model granularity and are generally considered as simplistic top-down approaches. Some country specific regulators adopted slightly modified versions of BIA and TSA.

Under the AMA, banks are allowed to use their own models to estimate the capital. A bank intending to use AMA should demonstrate the accuracy of the internal models within Basel II specified risk cells (8 business lines by 7 event types) relevant to the bank. This is a finer level of granularity more appropriate for a detailed analysis of risk processes in the financial institution. Typically, at this level of granularity,  the models are based on bottom-up approaches. The most widely used AMA is the loss distribution approach (LDA) based on modelling the annual frequency $N$ and severities $X_1,X_2,\ldots$ of OpRisk losses for a risk cell, so that the annual loss for a bank over the $d$ risk cells is
\begin{equation}
Z=\sum_{j=1}^d \sum_{i=1}^{N_j} X_i^{(j)}.
\end{equation}
Then, the regulatory capital is calculated as the 0.999 Value-at-Risk (VaR) which is the quantile of the distribution for the next year annual loss $Z$:
\begin{equation}
K_{LDA}=VaR_{q}[Z]:=\inf\{z\in \mathbb{R}:\Pr[Z>z]\le 1- q\},\quad q=0.999,
\end{equation}
that can be reduced by expected loss covered through internal provisions. Typically, frequency and severities within a risk cell are assumed to be independent.

For around ten years, the space of OpRisk has been evolving under this model based structure. A summary of the the Basel accords over this period of time (Basel II--Basel III) can be captured as follows:
\begin{itemize}
\item{ensuring that capital allocation is more risk sensitive;
}
\item{enhance disclosure requirements which would allow market participants to assess the capital adequacy of an institution;
}
\item{ensuring that credit risk, OpRisk and market risk are quantified based on data and formal techniques; and
}
\item{attempting to align economic and regulatory capital more closely to reduce the scope for regulatory arbitrage.
}
\end{itemize}
While the final Basel accord has at large addressed the regulatory arbitrage issue, there are still areas where regulatory capital requirements will diverge from the economic capital.

However, recently it was observed by studies performed by the BCBS and several local banking regulators that BIA and TSA do not correctly estimate the OpRisk capital, i.e. gross income as a proxy indicator for OpRisk exposure appeared  to be not a good assumption. Also, it appeared that capital under AMA is difficult to compare across banks due to a wide range of practices adopted by different banks.

So at this point two options are available to further refine and standardize OpRisk modelling practices, through refinement of the BIA and TSA, and more importantly convergence within internal modelling in the AMA framework. Or alternatively, sadly the option adopted by the current round of the Basel Committee consultations \cite{BCBSd3552016} in Pillar 1, to remove all internal modelling and modelling practice in OpRisk in favour of an overly simplified ``\emph{one size fits all}''  SMA model.

    This paper is structured as follows. Section \ref{SMAformula_sec} formally defines the Basel proposed SMA. The subsequent sections involve a collection of summary results and comments for studies performed on the proposed SMA model. \emph{Capital instability} and \emph{sensitivity} are studied in Section \ref{capitalInstability_sec}. Section \ref{RiskResponsivity_sec} discusses	\emph{reduction of risk responsivity and incentivized risk taking}.	\emph{Discarding key sources of OpRisk data} is discussed in Section \ref{DataSources_sec}.	\emph{Possibility of super-additive capital under SMA} is examined in Section \ref{superadditvity1_sec}. Section \ref{OpCarEstimation_sec} summarizes the Basel Committee procedure for estimation of OpCar model and underlying assumptions, and discusses the issues with this approach. The paper then concludes with suggestions in Section \ref{StandardisationAMA} relating to maintaining the AMA internal model framework with standardization recommendations that could be considered to unify internal modelling of OpRisk.
%%%%%%%%%%%%%%%%%%%%%%%%%%%%%%%%%%%%%%%%%%%%%%%%%%%%%%%%%%%%%%%%%%%%
\section{Basel Committee proposed SMA}\label{SMAformula_sec}
%\section{Simplification of OpRisk Models is not Necessarily Standardization}
%%%%%%%%%%%%%%%%%%%%%%%%%%%%%%%%%%%%%%%%%%%%%%%%%%%%%%%%%%%%%%%%%%%%
\noindent This section introduces the new simplifications that are being proposed by the Basel Committee for models in OpRisk, starting with a brief overview of how this process was initiated by the OpRisk Capital-at-Risk (OpCar) model proposed in  \cite{BCBSd2912014} and then finishing with the current version of this approach known as SMA proposed in \cite{BCBSd3552016}.

%First we discuss the initial round of proposed changes and then continue to a detailed discussion of the current round of changes in light of the feedback and revisions made post the first round proposal.

We begin with a clarification comment on the first round of proposal that is important conceptually to clarify for practitioners.  On page 1 of \cite{BCBSd2912014} it is stated that ``\textsl{... Despite an increase in the number and severity of operational risk events during and after the financial crisis, capital requirements for operational risk have remained stable or even fallen for the standardised approaches. This indicates that the existing set of simple approaches for operational risk - the Basic Indicator Approach (BIA) and the Standardised Approach (TSA), including its variant the Alternative Standardised Approach (ASA) - do not correctly estimate the operational risk capital requirements of a wide spectrum of banks.}''

We do agree that in general there are many cases where banks will be under-capitalized for large crisis events such as the one that hit in 2008. Therefore, with the benefit of hindsight, it is prudent to reconsider these simplified models and look for improvements and reformulations that can be achieved in the wake of new information post the 2008 crisis. In fact, we would argue this is sensible practice in model assessment and model criticism post new information regarding model suitability.

As we observed, the BIA and TSA make very simplistic assumptions regarding capital. Namely that the gross income can be used as an adequate proxy indicator for OpRisk exposure and furthermore that a banks' OpRisk exposure increases linearly in proportion to revenue. The consultative document \cite[page 1]{BCBSd2912014} also makes two relevant points that ``\textsl{... the existing approaches do not take into account the fact that the relationship between the size and the operational risk of  bank does not remain constant or that operational risk exposure increases with a bank's size in a non-linear fashion...}''.

Furthermore, neither BIA nor TSA approaches have been recalibrated since 2004.
%This is partially understandable due to the crisis, but the fact that a revision schedule is never provided with the regulations we believe translates an implicit regulation will to create universal rules that remain in force indefinitely.
We believe that this is a huge mistake and models and calibrations should be tested regularly, and each piece of regulation should come with its revision plan. As has been seen form experience, the model assumption have typically turned out to be invalid in a dynamically changing non-stationary risk management environment.

The two main objectives of the OpCar model proposed in \cite{BCBSd2912014} were stated to be:
\begin{itemize}
\item[(i)] refining the OpRisk proxy indicator by replacing $GI$ with a superior indicator; and
\item[(ii)] improving calibration of the regulatory coefficients based on the results of the quantitative analysis.
\end{itemize}

To achieve this the Basel Committee argued, based on practical  grounds, that the model developed should be sufficiently simple to be applied with ``\emph{comparability of outcomes in the framework}'' and ``\emph{simple enough to understand, not unduly burdensome to implement, should not have too many parameters for calculation by banks and it should not rely on banks' internal models}''. However, they also claim that such a new approach should ``\emph{exhibit enhanced risk sensitivity}'' relative to the $GI$ based frameworks.

Additionally, such a one-size fits all framework ``\emph{should be calibrated according to the OpRisk profile of a large number of banks of different size and business models}.'' We disagree with this motivation, as many banks in different jurisdictions and for different bank size and different bank practice may indeed come from different population level distributions. In other words, the  OpCar approach assumes all banks have a common population distribution from which their loss experience is drawn, and that this will be universal no matter what your business practice, your business volume or your jurisdiction of operation. Such an assumption surely may lead to a less risk sensitive framework with poorer insight into actual risk processes in a given bank than a properly designed model developed for a particular business volume, operating region and business practice.

The background and some further studies on the precursor OpCar framework, that was originally supposed to replace just the BIA and TSA methods, is provided in Section \ref{OpCarEstimation_sec}. We explain how the OpCar simplified framework was developed, explain the fact it is based on an LDA model and a regression structure, and how this model was estimated and developed. Along the way we provide some scientific criticism of several technical aspects of the estimation and approximations utilised.
It is important to still consider such aspects as this model is the precursor to the SMA formula.  That is, a single LDA is assumed for a bank and single loss approximation (SLA) is used to estimate the 0.999 quantile of the annual loss. Four different severity distributions were fitted to the data from many banks and Poisson distribution is assumed for the frequency. Then a non-linear regression is used to regress the obtained bank capital (across many banks) to a different combinations of explanatory variables from bank books to end up with the OpCar formula.

The currently proposed SMA for OpRisk capital in \cite{BCBSd3552016} is the main subject of our paper. However, we note that it is based on the OpCar formulation which itself is nothing more that an LDA model applied in an overly simplified fashion at the institution top level.

In \cite{BCBSd3552016} it was proposed to replace all existing BIA, TSA and AMA approaches with the SMA calculating OpRisk capital as a function of the so-called \emph{business indicator} ($BI$) and \emph{loss component} ($LC$). Specifically, denote $X_i(t)$ as the $i$-th loss and $N(t)$ as the number of losses in year $t$. Then, the SMA capital $K_{SMA}$ is defined as
\begin{equation}
K_{SMA}(BI,LC)=\left\{
\begin{array}{ll}
  BIC, & \mbox{if Bucket 1}, \\
  110+ (BIC-110)\ln\left(\exp(1)-1+\frac{LC}{BIC}\right), & \mbox{if Buckets 2-5}.
\end{array}
 \right.
\end{equation}
Here,
\begin{equation}\label{LC_eq}
LC=7\frac{1}{T}\sum_{t=1}^{T}\sum_{i=1}^{N(t)} X_i(t)+7\frac{1}{T}\sum_{t=1}^{T}\sum_{i=1}^{N(t)} X_i(t) 1_{\{X_i>10\}}+5\frac{1}{T}\sum_{t=1}^{T}\sum_{i=1}^{N(t)} X_i(t) 1_{\{X_i(t)>100\}},
\end{equation}
where $T=10$ years (or at least  5 years for banks that do not have 10 years of good quality loss data in the transition period). Buckets and $BIC$ (\emph{business indicator component}) are calculated as
\begin{equation}
BIC=\left\{
\begin{array}{ll}
  0.11\times BI, & \mbox{if}\;\; BI\le 1000,\;\mbox{Bucket 1},  \\
  110+0.15\times(BI-1000), & \mbox{if}\;\; 1000<BI\le 3000,\;\mbox{Bucket 2},\\
  410+0.19\times(BI-3000), & \mbox{if}\;\; 3000<BI\le 10000,\;\mbox{Bucket 3},\\
  1740+0.23\times(BI-10000), & \mbox{if}\;\; 10000<BI\le 30000,\;\mbox{Bucket 4},\\
  6340+0.29\times(BI-30000), & \mbox{if}\;\; BI>30000,\;\mbox{Bucket 5}.
\end{array}
 \right.
\end{equation}

$BI$ is defined as a sum of three components: interest, lease and dividend component; services component; and financial component. It is made up of almost the same P\&L items used for calculation of $GI$ but combined in a different way; for precise formula, see \cite{BCBSd3552016}. All amounts in the above formulas are in Euro million.

%%%%%%%%%%%%%%%%%%%%%%%%%%%%%%%%%%%%%%%%%%%%%%%%%%%%%%%%%%%%%%%%%%%%
%%%%%%%%%%%%%%%%%%%%%%%%%%%%%%%%%%%%%%%%%%%%%%%%%%%%%%%%%%%%%%%%%%%%
\section{SMA Introduces Capital Instability}\label{capitalInstability_sec}
%%%%%%%%%%%%%%%%%%%%%%%%%%%%%%%%%%%%%%%%%%%%%%%%%%%%%%%%%%%%%%%%%%%%
%%%%%%%%%%%%%%%%%%%%%%%%%%%%%%%%%%%%%%%%%%%%%%%%%%%%%%%%%%%%%%%%%%%%
\noindent In our analysis we observed that SMA fails to achieve the objective of capital stability. In this section we consider several examples to illustrate this feature. In most cases we show results for only Lognormal severity; other distribution types considered in OpCar model lead to similar or even more pronounced feature.

\subsection{Capital instability examples}
\noindent Consider a simple representative model for a bank's annual OpRisk loss process comprised of the aggregation of two generic loss processes, one high frequency with low severity loss amounts and the other corresponding to low frequency and high severity loss amounts given by Poisson-Gamma and Poisson-Lognormal models respectively. We set the business indicator $BI$ constant to Euro 2 billion at half way within the interval for Bucket 2 of the SMA, we kept the model parameters static over time and simulated a history of 1,000 years of loss data for three differently sized banks (small, medium and large) using different parameter settings for the loss models to characterize such banks. For a simple analysis we set a small bank corresponding to capital in the order of Euro 10's of million average annual loss, a medium bank in the order of Euro 100's of million average annual loss and a large bank was set to have in the order of Euro 1 billion average annual loss.  We then studied the variability that may arise in the capital under the SMA formulation, under the optimal scenario that models did not change, model parameters were not recalibrated and the business environment did not change significantly, in the sense that BI was kept constant. In this case we observe the core variation that arise just from loss history experience of  banks of the three different sizes over time.

\begin{figure}[!h]
\captionsetup{width=1.05\textwidth}
\begin{center}
\includegraphics[scale=0.49]{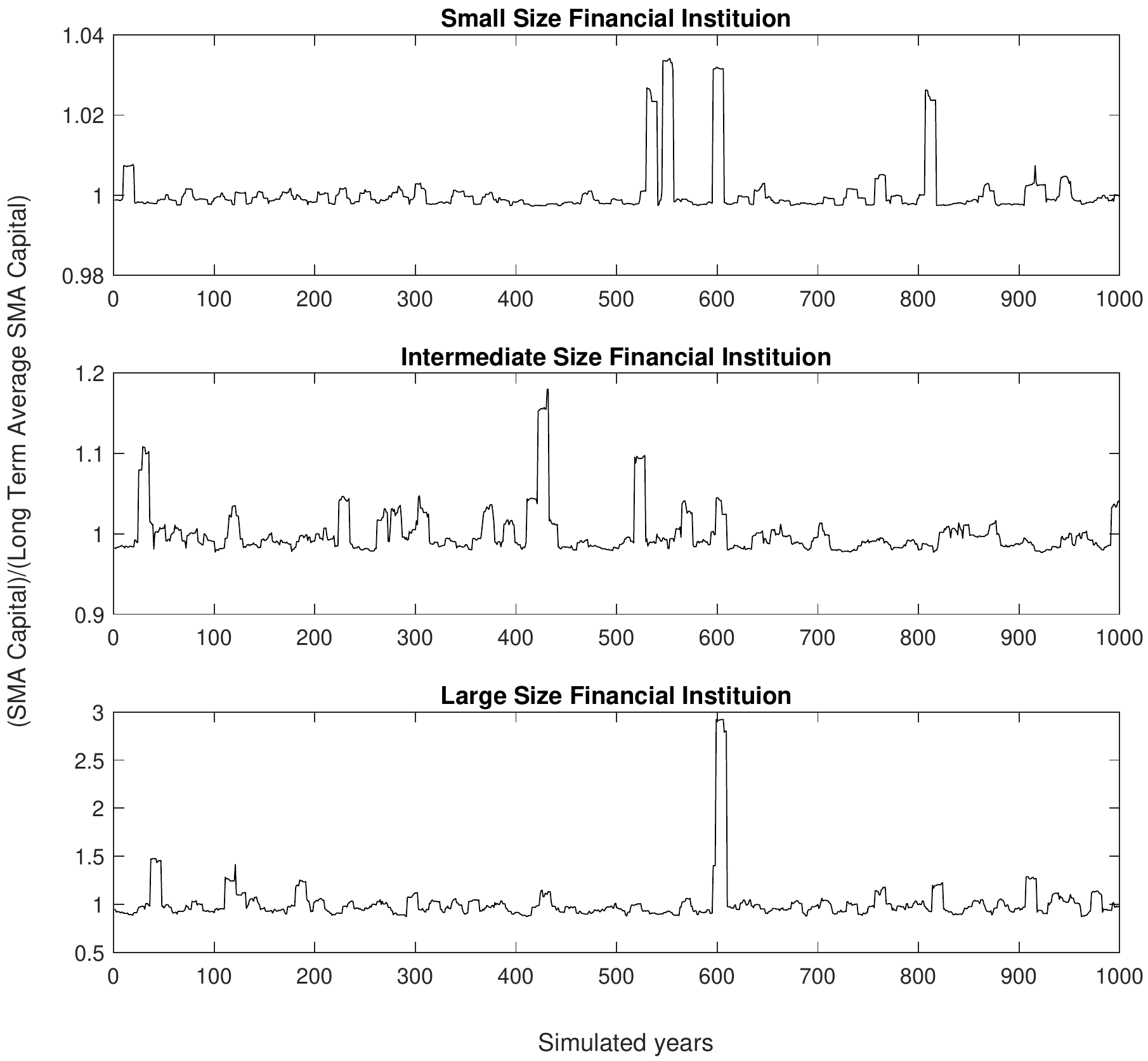}\includegraphics[scale=0.5]{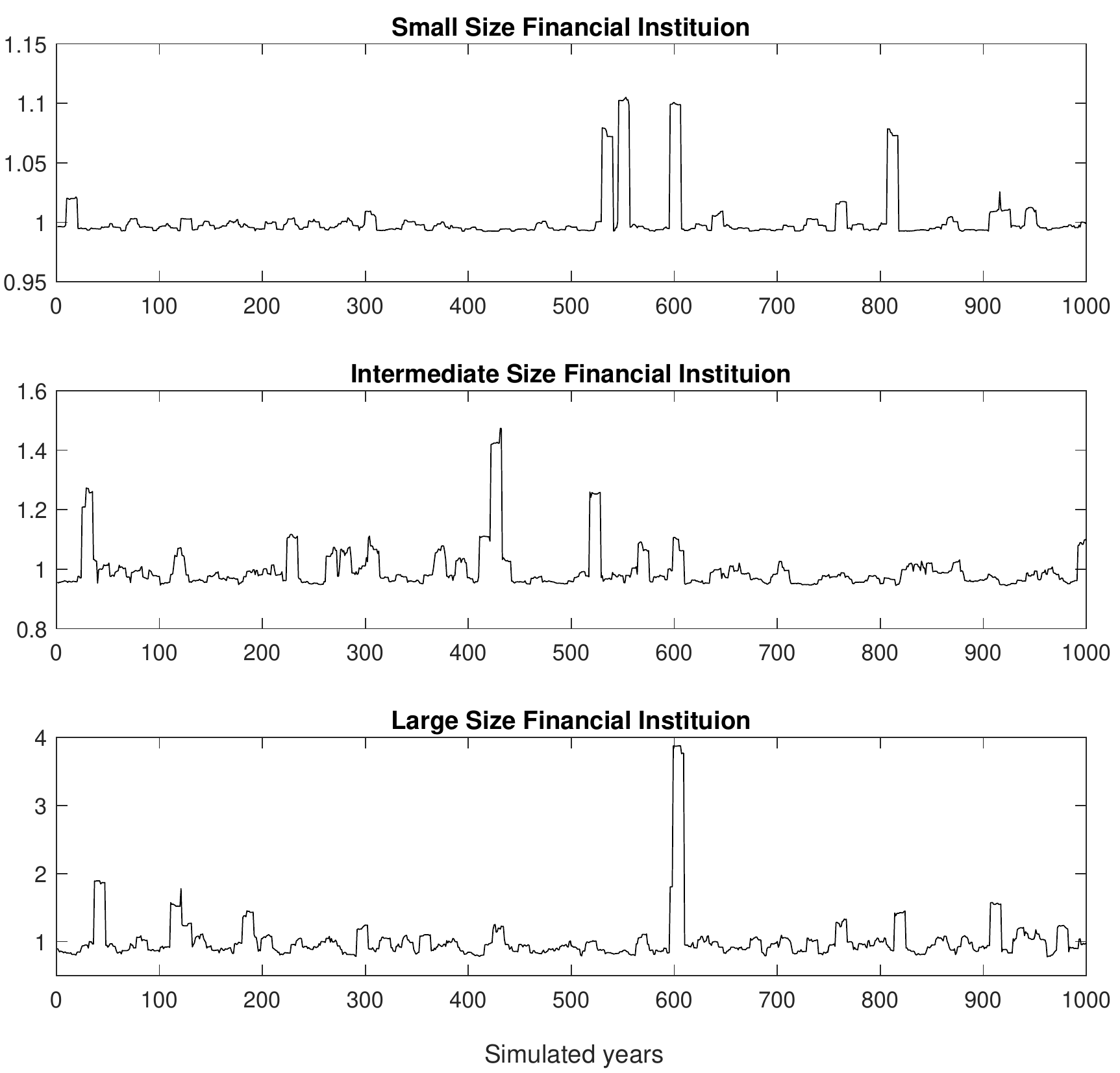}
\caption{\footnotesize{{Ratio of the SMA capital to the long term average. Test Case 1 corresponds to  $\sigma=2.5$ (plots on the left); Test Case 2 corresponds to $\sigma=2.8$ (plots on the right); other parameters are as specified in Table \ref{TestCases1and2_tab}. Results for small, intermediate and large banks are based on different realizations of random variables in simulation.}}}\label{TestCase1and2_fig}
\end{center}
\end{figure}

Our analysis shows that, a given institution can experience the situation in which its capital can more than double from one year to the next, without any changes to the parameters, the model or the BI structure (Figure \ref{TestCase1and2_fig}). %Annual variation can be as large as 2 times the long-term average capital.
This also means that two banks with the same risk profile can produce SMA capital numbers differing by a factor of more than 2.

 In summary, the simulation takes the case of $BI$ fixed over time, the loss model for the institution is fixed according to two independent loss processes given by $Poisson(\lambda)-Gamma(\alpha,\beta)$ and $Poisson(\lambda)-Lognormal(\mu,\sigma)$. Here, $Gamma(\alpha,\beta)$ is Gamma distribution of the loss severities with the mean $\alpha\beta$ and the variance $\alpha\beta^2$, and the $Lognormal(\mu,\sigma)$ is Lognormal distribution of severities with the mean of the log-severity equal to $\mu$ and the variance of the log-severity equal to $\sigma^2$.

The total institutions losses are set to be on average around 1,000 per year with 1\% coming from heavy tailed loss process Poisson-Lognormal component. We perform two case studies, one in which the shape parameter of the heavy tailed loss process component is $\sigma = 2.5$ and the other with $\sigma = 2.8$. We summarize the settings for the two cases below in Tables \ref{TestCases1and2_tab}.
The ideal situation that would indicate that SMA was not producing capital figures which were too volatile would be if each of the sub-figures below in Figure \ref{TestCase1and2_fig} was very closely constrained around 1. However, as we see, the variability in capital from year to year, in all size institutions can be very significant. Note that we used different sequences of independent random numbers to generate results for small, medium and large banks in a test case. Thus, a caution should be exercised in interpreting results of Test Case 1 (or Test Case 2) for relative comparison of capital variability of different banks. At the same time, comparing Test Case 1 with Test Case 2, one can certainly observe that increase in $\sigma$ increases the capital variability. %In particular medium to large size institutions both demonstrated that in any given year under SMA, the capital required to be held could double.
%These results demonstrate examples of typical variability in capital that can be experienced with the new SMA formulation.

\begin{table}[!h]\begin{center}
\captionsetup{width=0.95\textwidth}
\caption{\footnotesize{Test Case 1 corresponds to the risk process $Poisson(10)-Lognormal(\mu=\{10; 12; 14\},\sigma=2.5)$ and
$Poisson(990)-Gamma(\alpha=1,\beta =\{10^4;10^5; 5\times 10^5\})$. Test Case 2 corresponds to $Poisson(10)-Lognormal(\mu=\{10; 12; 14\},\sigma=2.8)$ and
$Poisson(990)-Gamma(\alpha=1,\beta =\{10^4;10^5; 5\times 10^5\})$.}} \label{TestCases1and2_tab}
{\footnotesize{
\begin{tabular*}{1.0\textwidth}{c|cc|cc}
\toprule
\multirow{3}{*}{bank size} & Test Case 1 & Test Case 1 & Test Case 2 & Test Case 2\\
&  Mean Annual Loss  &   Annual Loss 99.9\% VaR &  Mean Annual Loss  &   Annual Loss 99.9\% VaR\\
& (Euro Million) & (Euro Million) & (Euro Million) & (Euro Million)\\
 \midrule
small & 15 & 260& 21 & 772 \\
medium & 136& 1,841 & 181 & 5,457 \\
large & 769&14,610 & 1,101 &41,975 \\
\bottomrule
\end{tabular*}
}}
\end{center}
\end{table}
%\begin{table}[!h]\begin{center}
%\captionsetup{width=0.9\textwidth}
%\caption{Test Case 2 with risk process $Poisson(10)-Lognormal(\mu=\{10; 12; 14\},\sigma=2.8)$ and
%$Poisson(990)-Gamma(\alpha=1,\beta =\{10^4;10^5; 5\times 10^5\})$.} \label{tab2}
%{\footnotesize{
%\begin{tabular*}{0.55\textwidth}{ccc}
%\toprule
%\multirow{2}{*}{bank size} &  Mean Annual Loss  &   Annual Loss 99.9\% VaR\\
%& (Euro Million) & (Euro Million)\\
% \midrule
%small & 21 & 772 \\
%medium & 181 & 5,457 \\
%large & 1,101 &41,975 \\
%\bottomrule
%\end{tabular*}
%}}
%\end{center}
%\end{table}
%\begin{figure}[!h]
%\captionsetup{width=0.9\textwidth}
%\begin{center}
%\includegraphics[scale=0.6]{Fig2}
%\caption{{{Ratio of the SMA capital to the long term average (Case 2). Note: results in each subplot utilised different realizations of random variables in Monte Carlo simulation.}}}\label{fig2}
%\end{center}
%\end{figure}

%%%%%%%%%%%%%%%%%%%%%%%%%%%%%%%%%%%%%%%%%%%%%%%%%%%%%%%%%%%%%%%%%%%%
%%%%%%%%%%%%%%%%%%%%%%%%%%%%%%%%%%%%%%%%%%%%%%%%%%%%%%%%%%%%%%%%%%%%
\subsection{Capital instability and BI when SMA matches AMA}\label{ImpliedBI_capital_instability}
%%%%%%%%%%%%%%%%%%%%%%%%%%%%%%%%%%%%%%%%%%%%%%%%%%%%%%%%%%%%%%%%%%%%
%%%%%%%%%%%%%%%%%%%%%%%%%%%%%%%%%%%%%%%%%%%%%%%%%%%%%%%%%%%%%%%%%%%%
\noindent As a second study of the SMA capital instability we  consider a loss process model $Poisson(\lambda)-Lognormal(\mu,\sigma)$. Instead of fixing the $BI$ to the midpoint of Bucket 2 of the SMA formulation, we numerically solve for the $BI$ that would produce the SMA capital equal to the Value-at-Risk for a Poisson-Lognormal LDA model at the annual 99.9\% quantile level, $VaR_{0.999}$.

In other words,  we find the $BI$ such that the LDA capital matches SMA capital in the long term. This is achieved by solving the following non-linear equation numerically via root search for the $BI$:
\begin{equation}\label{ImpliedBI_eq}
K_{SMA}(BI,\widetilde{LC}) = VaR_{0.999},
\end{equation}
where $\widetilde{LC}$ is the long term average of the loss component (\ref{LC_eq}) that can be calculated in the case of $Poisson(\lambda)$ frequency as
\begin{equation}\label{LTALC_eq}
\widetilde{LC}=\lambda\times\bigg(7\, \mathrm{E}[X]+7\, \mathrm{E}[X|X>L]+5\, \mathrm{E}[X|X>H]\bigg).
\end{equation}
In the case of severity $X$ from $Lognormal(\mu,\sigma)$, it can be found in closed form as
\begin{equation}\label{LTALC_LN_eq}
\widetilde{LC}(\lambda,\mu,\sigma)=\lambda e^{\mu+\frac{1}{2}\sigma^2}\left(7+7\, \Phi\left(\frac{\sigma^2+\mu-\ln L}{\sigma}\right)+5\,\Phi\left(\frac{\sigma^2+\mu-\ln H}{\sigma}\right) \right),
\end{equation}
where $\Phi(\cdot)$   denotes the standard Normal distribution function, $L$ is Euro 10 million and $H$ is Euro 100 million as specified by SMA formula (\ref{LC_eq}).

One can approximate the $VaR_{0.999}$ under the Poisson-Lognormal model according to the so-called single loss approximation (discussed in Section \ref{OpCarEstimation_sec}), given for $\alpha \uparrow 1$ by
\begin{equation} \label{SLALN}
\begin{split}
VaR_{\alpha}&\approx SLA(\alpha;\lambda,\mu,\sigma) \\
& = \exp\left(\mu+\sigma\Phi^{-1}\left(1-\frac{1-\alpha}{\lambda}\right)\right)+\lambda\exp\left(\mu+\frac{1}{2}\sigma^2 \right),
\end{split}
\end{equation}
where $\Phi^{-1}(\cdot)$  is the inverse of the standard Normal distribution function. In this case, the results for the implied $BI$ values are presented in Table \ref{ImpliedBI_tab} for $\lambda = 10$ and varied Lognormal $\sigma$ and $\mu$ parameters. Note that it is also not difficult to calculate $VaR_{\alpha}$ ``exactly" (within numerical error) using Monte Carlo, Panjer recursion or FFT numerical methods.

\begin{table}[!h]\begin{center}
\captionsetup{width=0.6\textwidth}
\caption{\footnotesize{{Implied $BI$ in billions, $\lambda=10$.}}} \label{ImpliedBI_tab}
{\footnotesize{
\begin{tabular*}{0.65\textwidth}{c|ccccccc} \toprule
$\mu\setminus\sigma$ & 1.5 &  1.75 &   2.0 & 2.25 & 2.5 & 2.75 & 3.0 \\
 \midrule
10	     & 0.06	     & 0.14	     & 0.36	     & 0.89	     & 2.41	     & 5.73	     & 13.24\\
12		&0.44		&1.05		&2.61		&6.12		&14.24		&32.81		&72.21\\
14		&2.52		&5.75		&13.96		&33.50		&76.63		&189.22		&479.80\\
\bottomrule
\end{tabular*}
}}
\end{center}
\end{table}

\begin{figure}[!h]
\captionsetup{width=0.9\textwidth}
\begin{center}
\includegraphics[scale=0.7]{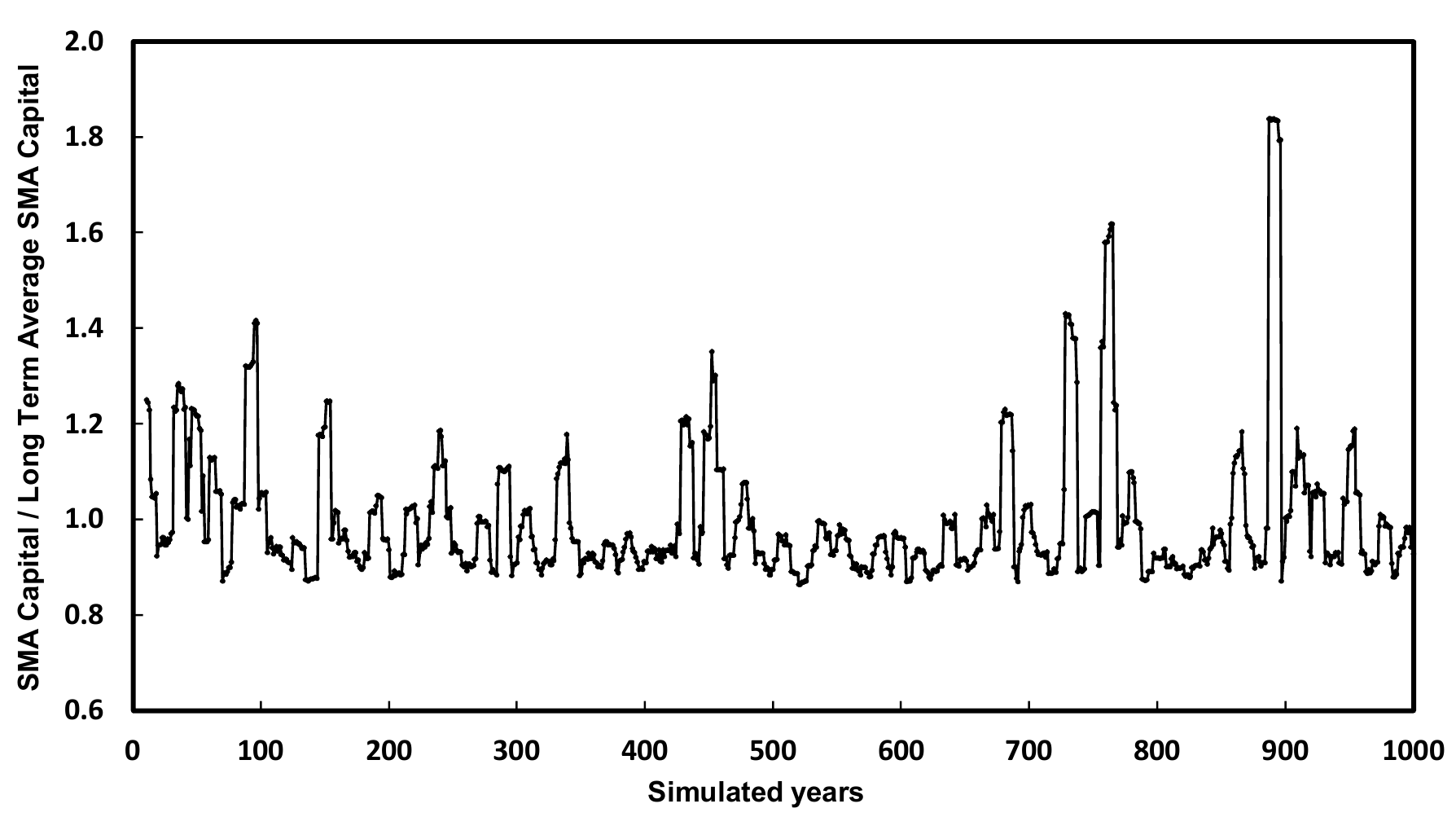}
\caption{\footnotesize{{Ratio of the SMA capital to the long term average. }}}\label{CapitalStability_ImpliedBI_fig}
\end{center}
\end{figure}

For a study of capital instability, we use the $BI$ obtained from matching the long term average SMA capital with the long term LDA capital, as described above for an example generated by $Poisson(10)-Lognormal(\mu =12, \sigma =2.5)$ and correspondingly found implied $BI=$ Euro 14.714 billion (Bucket 4). In this case we calculate  $VaR_{0.999}$ using Monte Carlo instead of a single loss approximation (\ref{SLALN}) and thus the value of implied $BI$ is slightly different from the one in Table \ref{ImpliedBI_tab}.  In this case the SMA capital based on the long term average $LC$ is Euro 1.87 billion is about the same as $VaR_{0.999}$ = Euro 1.87 billion. Then the year on year variability in the capital with this combination of implied $BI$ and Poisson-Lognormal loss model is given in Figure \ref{CapitalStability_ImpliedBI_fig}. It shows that again we get capital instability with capital doubling from year to year compared to the long term average SMA capital.

\FloatBarrier
%%%%%%%%%%%%%%%%%%%%%%%%%%%%%%%%%%%%%%%%%%%%%%%%%%%%%%%%%%%%%%%%%%%%
%%%%%%%%%%%%%%%%%%%%%%%%%%%%%%%%%%%%%%%%%%%%%%%%%%%%%%%%%%%%%%%%%%%%
\subsection{SMA is Excessively Sensitive to the Dominant Loss Process}
\begin{figure}[!h]
\captionsetup{width=0.9\textwidth}
\begin{center}
\includegraphics[scale=0.6]{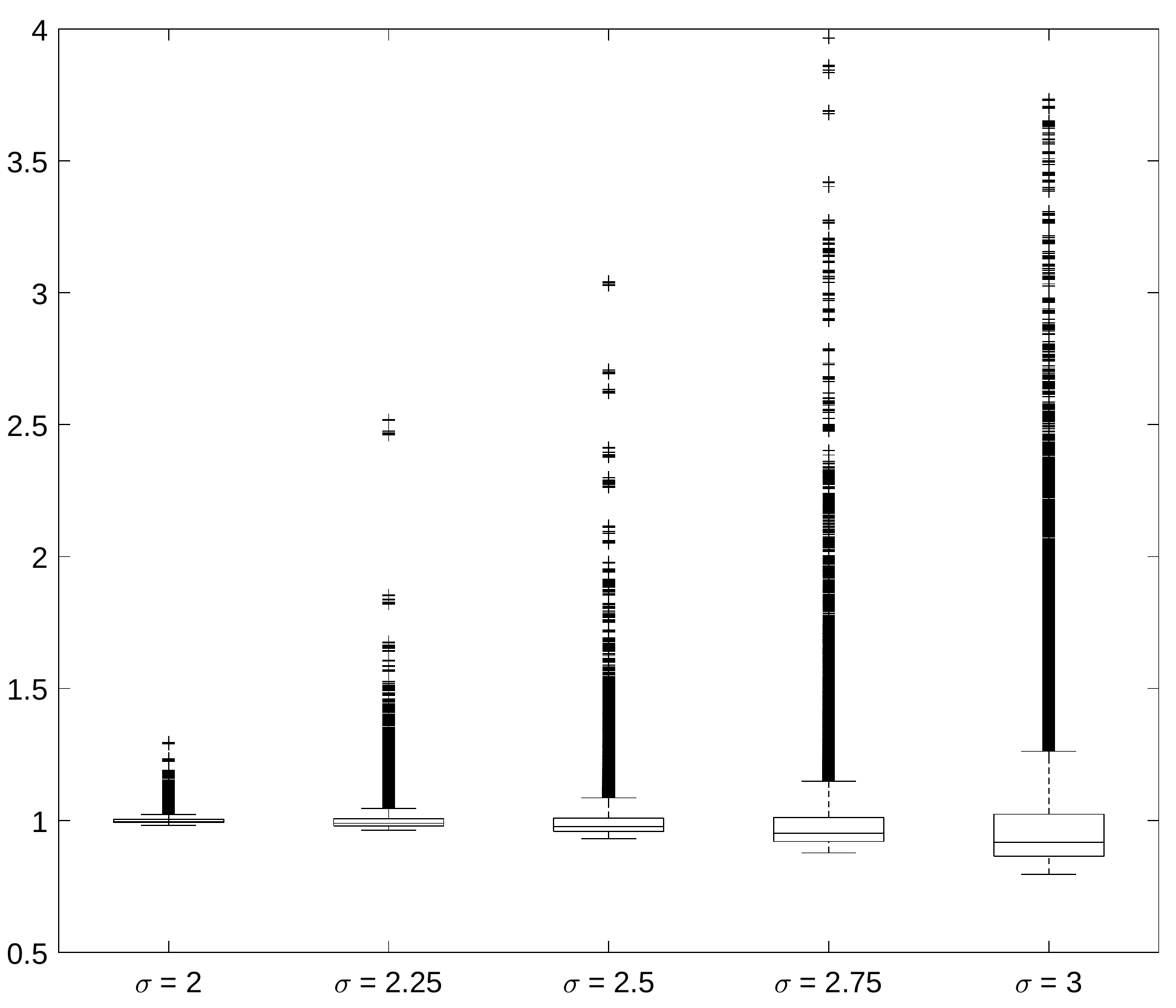}
\caption{\footnotesize{{Boxplot results for the ratio of the SMA capital to the long term average for different values of the Lognormal shape parameter $\sigma$ based on simulations over 1,000 years. }}}\label{fig4}
\end{center}
\end{figure}

%%%%%%%%%%%%%%%%%%%%%%%%%%%%%%%%%%%%%%%%%%%%%%%%%%%%%%%%%%%%%%%%%%%%
%%%%%%%%%%%%%%%%%%%%%%%%%%%%%%%%%%%%%%%%%%%%%%%%%%%%%%%%%%%%%%%%%%%%
\noindent Consider an institution with a wide range of different types of OpRisk loss processes present in each of its business units and risk types. As in our first study above, we split these loss processes into two categories: high-frequency/low-severity, and low-frequency/high-severity types, given by $Poisson(990)-Gamma(1,5\times 10^5)$ and $Poisson(10)-Lognormal(14,\sigma)$ respectively. In this study we consider the sensitivity of SMA capital to the dominant loss process. More precisely, we study the sensitivity of SMA capital to the parameter $\sigma$ that dictates how heavy the tail of the most extreme loss process will be. Figure \ref{fig4} shows boxplot results based on simulations performed over 1,000 years for different values $\sigma=\{2; 2.25; 2.5; 2.75; 3\}$.

%\begin{table}[!h]\begin{center}
%\captionsetup{width=0.5\textwidth}
%\caption{Test Case 3, varying $\sigma$.} \label{tab4}
%{\footnotesize{
%\begin{tabular*}{0.6\textwidth}{ll} \toprule
%Model & Parameters \\
% \midrule
%Poisson-Lognormal 	     & $(\lambda,\sigma, \mu) = (10, \{2; 2.25; 2.5; 2.75; 3\} , 14)$\\
%Poisson-Gamma 		& $(\lambda,\alpha,\beta) =  (990, 1, 500,000)$\\
%\bottomrule
%\end{tabular*}
%}}
%\end{center}
%\end{table}

These results can be interpreted to mean that banks with more extreme loss experiences as indicated by heavier tailed dominant loss processes (increasing $\sigma$) tend to have significantly greater capital instability compared to banks with less extreme loss experiences. Importantly, these findings demonstrate how non-linear this increase in SMA capital can be as the heaviness of the dominant loss process tail increases. For instance, banks with relatively low heavy tailed dominant loss processes ($\sigma=2$) tend to have capital variability year on year of between 1.1 to 1.4 multipliers of long term average SMA capital. However, banks with relatively large heavy tailed dominant loss processes ($\sigma$ = 2.5, 2.75 or 3) tend to have excessively unstable year on year capital figures, with variation in capital being as bad as 3 to 4 times multipliers of the long term average SMA capital. Furthermore, it is clear that when one considers each boxplot as representing a population of banks with similar dominant loss process characteristics, the population distribution of capital becomes increasingly skewed and demonstrates increasing kurtosis in the right tail as the tail heaviness of the dominant loss process in each population increases.
 This clearly demonstrates excessive variability in capital year on year for banks with heavy tailed dominant loss processes.

Therefore, SMA fails to achieve the claimed objective of robust capital estimation.  Capital produced by the proposed SMA approach will be neither stable nor robust with worsening robustness as the severity of OpRisk increases. In other words, banks with higher severity OpRisk exposures will be substantially worse of under the SMA approach with regard to capital sensitivity.

%%%%%%%%%%%%%%%%%%%%%%%%%%%%%%%%%%%%%%%%%%%%%%%%%%%%%%%%%%%%%%%%%%%%
%%%%%%%%%%%%%%%%%%%%%%%%%%%%%%%%%%%%%%%%%%%%%%%%%%%%%%%%%%%%%%%%%%%%
\section{Reduced Risk Responsivity and Induced Risk-Taking }\label{RiskResponsivity_sec}
 It is obvious that the SMA capital is less responsive to risk drivers and the variation in loss experience that is observed in a bank at granularity of the Basel II 56 business line/event type units of measure.

This is due to the naive approach of modelling at the level of granularity assumed by the SMA which is only capturing variability at the institution level and not the intra-variability within the institution at business unit levels explicitly. Choosing to model at institution level, rather than the units of measure or granularity of the 56 Basel categories reduces model interpretability and reduces risk responsivity of the capital.

Conceptually, it relates to the simplification of the AMA under the SMA adopting a top down formulation that reduces OpRisk modelling to a single unit of measure, as if all operational losses were following a single generating mechanism. This is equivalent to considering that earthquakes, cyber-attacks and human errors are all generated by the same drivers and manifest in the loss model and loss history in the same manner as other losses that are much more frequent and have lower consequence, such as credit card fraud, when viewed from the institution level loss experience. It follows quite obviously that the radical simplification and aggregation of such heterogeneous risks in such a simplified model cannot claim the benefit of risk-sensitivity, even remotely.

Therefore, SMA fails to achieve the claimed objective of capital risk sensitivity. Capital produced by the proposed SMA approach will be neither stable nor related to the risk profile of an institution.
%%%%%%%%%%%%%%%%%%%%%%%%%%%%%%%%%%%%%%%%%%%%%%%%%%%%%%%%%%%%%%%%%%%%
%%%%%%%%%%%%%%%%%%%%%%%%%%%%%%%%%%%%%%%%%%%%%%%%%%%%%%%%%%%%%%%%%%%%
%\section{SMA Incentivizes Enhanced Risk-Taking}\label{IncentiveRiskTaking_sec}
%%%%%%%%%%%%%%%%%%%%%%%%%%%%%%%%%%%%%%%%%%%%%%%%%%%%%%%%%%%%%%%%%%%%
%%%%%%%%%%%%%%%%%%%%%%%%%%%%%%%%%%%%%%%%%%%%%%%%%%%%%%%%%%%%%%%%%%%%
%\noindent
Moreover, the SMA induces risk-taking behaviors, failing to achieve the Basel committee objectives of stability and soundness of the financial institutions.
Moral hazard and other unintended consequences include as follows.

\vspace{0.3cm}

\noindent{\textbf{More risk-taking.}} Without the possibility of capital reduction for better risk management, in the face of increased funding costs due to the rise in capital, it is predictable that financial institutions will raise their risk-taking to a level sufficient enough to pay for the increased cost of the new fixed capital. The risk appetite a financial institution would mechanically increase. This effect goes against the Basel Committee objective of a safe and secured financial system.

\vspace{0.3cm}

\noindent{\textbf{Denying loss events.}} Whilst incident data collection is a constant effort for over a decade in every institutions, large or small, the SMA is the most formidable disincentive to report losses. There are many opportunities to compress historical losses such as ignoring, slicing, transferring to other risk categories. The wish expressed in the Basel consultation that ``\emph{Banks should use 10 years of good-quality loss data}" is actually meaningless if the collection can be gamed. Besides, what about new banks or BIA banks which do not have any loss data collection process as of today?

\vspace{0.3cm}

\noindent{\textbf{Hazard of reduced provisioning activity.}} Provisions, which should be a substitution for capital, are vastly discouraged by the SMA, as they are penalized twice, counted both in the BI and in the losses, and not accounted for as a capital reduction. The SMA captures both the expected loss and the unexpected loss, when the regulatory capital should only reflect the unexpected loss. We believe that this confusion might come from the use of the OpCar model as a benchmark because the OpCar captures both equally.
The SMA states in the definition of gross loss, net loss, and recovery on page 10 section 6.2 of \cite{BCBSd3552016} under item (c) that the gross loss and net loss should include ``\emph{Provisions or reserves accounted for in the P\&L against the potential operational loss impact}". This clearly indicates the nature of the double counting of this component since they will enter in the BI through the P\&L and in the loss data component of the SMA capital.

\vspace{0.3cm}

\noindent{\textbf{Ambiguity in provisioning and resulting capital variability.}} The new guidelines on provisioning under SMA framework follow similar general concept as those that recently came into effect in credit risk with the International Financial Reporting Standard (IFRS9) set forward by the International Accounting Standards Board (IASB), who completed the final element of its comprehensive response to the financial crisis with the publication of IFRS9 Financial Instruments in July 2014. The IFRS9 guidelines explicitly outline in Phase 2 an impairment framework which specifies in a standardized manner how to deal with delayed recognition of (in this case) credit losses on loans (and other financial instruments).
IFRS9 achieves this through a new expected loss impairment model that will require more timely recognition of expected credit losses. Specifically, the new Standard requires entities to account for expected credit losses from when financial instruments are first recognised and it lowers the threshold for recognition of full lifetime expected losses.
However, the SMA OpRisk version of such provisioning concept for OpRisk losses fails to provide such a standardized and rigorous approach. Instead, the SMA framework simply states that loss databases should now include
``\emph{Losses stemming from operational risk events with a definitive financial impact, which are temporarily booked in transitory and/or suspense accounts and are not yet reflected in the P\&L (``pending losses''). Material pending losses should be included in the SMA loss data set within a time period commensurate with the size and age of the pending item}''.
Unlike the more specific IFRS9 accounting standards, under the SMA there is a level of ambiguity. Furthermore, this ambiguity can propagate now directly into the SMA capital calculation causing potential for capital variability and instability.
For instance, there is no specific guidance or regulation requirements to standardize the manner in which a financial institution decides what is to be considered as ``definitive financial impact'' and what they should consider as a threshold for deciding on existence of a ``material pending loss''. Also, it is not stated what is specifically the guidance or rule about the time periods related to inclusion of such pending losses in an SMA loss dataset and therefore into the capital. The current guidance simply states ``\emph{Material pending losses should be included in the SMA loss dataset within a time period commensurate with the size and age of the pending item}''. This is too imprecise and may lead to manipulation of provisions reporting and categorization that will directly reduce the SMA capital over the averaged time periods in which the loss data component is considered.
 Furthermore, if different financial institutions adopt differing provisioning rules, the capital obtained for two banks with identical risk appetites and similar loss experience could differ substantially just as a result of their provisioning practices.

\vspace{0.3cm}

\noindent{\textbf{Imprecise Guidance on Timing Loss Provisioning.}}
The SMA guidelines also introduce an aspect of ``Timing Loss Provisioning" in which they state:
``\emph{Negative economic impacts booked in a financial accounting period, due to operational risk events impacting the cash flows or financial statements of previous financial accounting periods (timing losses). Material ``timing losses'' should be included in the SMA loss data set when they are due to operational risk events that span more than one financial accounting period and give rise to legal risk}."
However, we would argue that for standardization of a framework there needs to be more explicit guidance as to what constitutes a ``Material timing loss". Otherwise, different timing loss provisioning approaches will result in different loss databases and consequently can result in differing SMA capital just as a consequence of the provisioning practice adopted. In addition, the ambiguity of this statement does not make it clear as to whether such losses may be accounted for twice.

\vspace{0.3cm}

\noindent{\textbf{Grouping of Losses.}} Under previous AMA internal modelling approaches the unit of measurement or granularity of the loss modelling was reported according to the 56 business line/event type categories specified in Basel II framework. However, under the SMA the unit of measure is just at the institution level so the granularity of the loss processes modelling and interpretation is lost. This has consequences when it is considered in light of the new SMA requirement that
``\emph{Losses caused by a common operational risk event or by related operational risk events over time must be grouped and entered into the SMA loss data set as a single loss}."
Previously, in internal modelling losses within a given business line/event type would be recorded as a random number (frequency model) of individual independent loss amounts (severity model). Then, for instance under an LDA model such losses would be aggregated only as a compound process and the individual losses would not be ``grouped" except on the annual basis and not on the per-event basis. However, there seems to be a marked difference in the SMA loss data reporting on this point, under the SMA it is proposed to aggregate the individual losses and report them in the loss database as a ``single grouped" loss amount. This is not advisable from a modelling or an interpretation and practical risk management perspective.
Furthermore, the SMA guidance states
``\emph{The bank's internal loss data policy should establish criteria for deciding the circumstances, types of data and methodology for grouping data as appropriate for its business, risk management and SMA regulatory capital calculation needs}."
One could argue that if the aim of SMA was to standardize OpRisk loss modelling in order to make capital less variable due to internal modelling decisions, then one can fail to see how this will be achieved with imprecise guidance such as the one provided above. One could argue that the above generic statement on criteria establishment basically removes the internal modelling framework of AMA and replaces it with internal heuristic (non-model based, non-scientifically verifiable) rules to ``group" data. This has the potential to result in even greater variability in capital than was experienced with non-standardized AMA internal models. At least under AMA internal modelling, in principle, the statistical models could be scientifically criticized.

\vspace{0.3cm}

\noindent{\textbf{Ignoring the future.}} All forward-looking aspects of risk identification, assessment and mitigation such as scenarios and emerging risks have disappeared in the new Basel consultation. This in effect introduces the risk of setting back the banking institutions in their progress towards a better understanding of the threats, even though such threats may be increasing in frequency and severity and the bank exposure to such threats may be increasing due to business practices, this cannot be reflected in the SMA framework capital. In that sense, the SMA is only backward looking.

%%%%%%%%%%%%%%%%%%%%%%%%%%%%%%%%%%%%%%%%%%%%%%%%%%%%%%%%%%%%%%%%%%%%
%%%%%%%%%%%%%%%%%%%%%%%%%%%%%%%%%%%%%%%%%%%%%%%%%%%%%%%%%%%%%%%%%%%%
\section{SMA Fails to Utilize Range of Data Sources and Fails to Provide Risk Management Insight }\label{DataSources_sec}
%%%%%%%%%%%%%%%%%%%%%%%%%%%%%%%%%%%%%%%%%%%%%%%%%%%%%%%%%%%%%%%%%%%%
%%%%%%%%%%%%%%%%%%%%%%%%%%%%%%%%%%%%%%%%%%%%%%%%%%%%%%%%%%%%%%%%%%%%
\noindent As with any scientific discipline, OpRisk modelling is no different when it comes to developing a statistical modelling framework. In practical settings it is therefore important to set the context with respect to the data and the regulatory requirements of Basel II with regard to data used in OpRisk modelling. The data aspect of OpRisk modelling has been an ongoing challenge for banks to develop suitable loss databases to record observed losses internally and externally along side other important information that may aid in modelling.

A key process in OpRisk modelling has not just been the collection but importantly how and what to collect as well as how to classify. A first and key phase in any analytical process, certainly the case in OpRisk models, is to cast the data into a form amenable to analysis. This is the very first task that an analyst faces when determined to model, measure and even manage OpRisk. At this stage there is a need to establish how the information available can be modelled to act as an input in the analytical process that would allow proper risk assessment processes to be developed.  In risk management, and particularly in OpRisk, this activity is today quite regulated and the entire data process, from collection to maintenance and use has strict rules. In this sense we see that qualitative and quantitative aspects of OpRisk management cannot be dissociated as they act in a causal manner on one another.

Any OpRisk modelling framework starts by having solid risk taxonomy so risks are properly classified. Firms also need to perform a comprehensive risk mapping across their processes to make sure that no risk is left out of the measurement process. This risk mapping is particularly important as it directly affects the granularity of the modelling, the features observed in the data, the ability to interpret the loss model outputs as well as the ability to collect and model data. Furthermore, it can affect the risk sensitivity of the models. It is a process that all large financial institutions have gone through at great cost, man power and time in order to meet compliance with Basel II regulations.

Under the Basel II regulations there are four major data elements that should be used to measure and manage
OpRisk: \textbf{internal loss data}; \textbf{external loss data}; \textbf{scenario analysis}; and \textbf{business environment and internal control factors (BEICFs)}.

To ensure that data is correctly specified in an OpRisk modelling framework, one must undertake a risk mapping or taxonomy exercise which basically encompasses: description; identification; nomenclature; and classification. This is a very lengthy and time consuming process that has typically been done by many banks at a fairly fine level of granularity with regard to the business structure. It involves going through, in excruciating details, every major process of the firm. The outcome of this exercise would be the building block of any risk classification study. We also observe that in practice, this task is complicated by the fact that in OpRisk settings, often when a risk materialises, and until it is closed, the loss process will continue to evolve over time, sometimes for many years if we consider legal cases. And in some cases, the same list of incidents taken at two different time points will not have the same distribution of loss magnitude. Here, it is important to bear in mind that a risk is not a loss, we may have risk and never experience an incident, and we may have incidents and never experience a loss. These considerations should also be made when developing a classification or risk mapping process.

There are roughly three ways that firms drive this risk taxonomy exercise: \emph{cause-driven; impact-driven; and event-driven}. The event-driven risk classification is probably the most common one used by large firms and has been the emerging best practice in OpRisk. This process classifies risk according to OpRisk events. This is the classification used by the Basel Committee for which a detailed breakdown into event types at level 1, level 2 and activity groups is provided in \cite[pp. 305-307]{BaselII2006}. Furthermore, it is generally accepted that this classification has a definition broad enough to make it easier to accept/adopt changes in the process should they arise. Besides, it is very interesting to note that a control taxonomy may impact the perception of events in the risk taxonomy, especially if the difference between inherent and residual risk is not perfectly understood. The residual risks are defined as \textsl{inherent risk-controls}, i.e. once we have controls in place we manage the residual risks while we might still be reporting inherent risks and this may bias the perception of the bank risk profile. The risk/control relationship (in terms of taxonomy) is not that easy to handle as risk owners and control owners might be in completely different departments, preventing a smooth transmission of information. This we believe also needs further consideration in emerging best practice and governance implications for OpRisk management best practice.

%%%%%%%%%%%%%%%%%%%%%%%%%%%%%%%%%%%%%%%%%%%%%%%%%%%%%%%%%%%%%%%%%%%%%%%%%%%%%%
\subsection{The Elements of the OpRisk Framework}
%%%%%%%%%%%%%%%%%%%%%%%%%%%%%%%%%%%%%%%%%%%%%%%%%%%%%%%%%%%%%%%%%%%%%%%%%%%%%%
\noindent The four key elements that should be used in any OpRisk framework are:
\emph{internal loss data; external loss data; BEICFs; and scenario analysis}.

In terms of OpRisk losses, typically, the definition means a gross monetary loss or a net monetary loss, i.e. net of recoveries but excluding insurance or tax effects, resulting from an operational loss event. An operational loss includes all expenses associated with an operational loss event except for opportunity costs, forgone revenue, and costs related to risk management and control enhancements implemented to prevent future operational losses. These losses need to be classified into the Basel categories (and internal if different than the Basel) and mapped to a firm's business units. Basel II regulation says that firms need to collect at least five years of data, but most decided not to discard any loss even when these are older than this limit. Losses are difficult to acquire and most even pay to supplement internal losses with external loss databases. Considerable challenges exist in collating a large volume of data, in different formats and from different geographical locations, into a central repository, and ensuring that these data feeds are secure and can be backed up and replicated in case of an accident.

%\begin{remark}
%Banks usually use the largest horizon possible despite the fact that sometimes they are using outdated losses that might positively or negatively bias the capital calculation through an inappropriate capture of the risk profile.
%\end{remark}

There is also a considerable challenge with OpRisk loss data recording and reporting related to the length for resolution of OpRisk losses. Some OpRisk events, usually some of the largest, will have a large time gap between the inception of the event and final closure, due to the complexity of these cases. As an example, most litigation cases that came up from the financial crisis in 2007/8 were only settled by 2012/13. These legal cases have their own life cycle and start with a discovery phase in which layers and investigators would argue if the other party have a proper case to actually take the action to court or not. At this stage it is difficult to even come up with an estimate for eventual losses. Even when a case is accepted by the judge it might be several years until lawyers and risk managers are able to estimate properly the losses.

Firms can set up reserves for these losses (and these reserves should be included in the loss database) but they usually only do that a few weeks before the case is settled to avoid disclosure issues (i.e. the counterparty eventually knowing the amount reserved and use this information in their favor). This creates an issue for setting up OpRisk capital because firms would know that a large loss is coming but they cannot include it yet in the database, the inclusion of this settlement would cause some volatility in the capital. The same would happen if a firm set a reserve of, for example, USD 1 billion for a case and then a few months later a judge decide in the firm's favor and this large loss has to be removed. For this reason, firms need to have a clear procedure on how to handle those large, long duration losses.

The other issue with OpRisk loss reporting and recording is the aspect of adding costs to losses. As mentioned, an operational loss includes all expenses associated with an operational loss event except for opportunity costs, forgone revenue, and costs related to risk management and control enhancements implemented to prevent future operational losses. Most firms, for example, do not have enough lawyers on payroll (or expertise) to deal with all the cases, particularly some of the largest or those that demand some specific expertise and whose legal fees are quite expensive. There will be cases in which the firm wins in the end, maybe due to the external law firms, but the cost can reach tens of millions of dollars. In this case, even with a court victory, there will be an operational loss. This leads to the consideration of provisioning of expected OpRisk losses. Unlike credit risk when the calculated expected credit losses might be covered by general and/or specific provisions in the balance sheet. For OpRisk, due to its multidimensional nature, the treatment of expected losses is more complex and restrictive. Recently, with the issuing of IAS37 by the International Accounting Standards Board IFRS2012, the rules have become clearer as to what might be subject to provisions (or not). IAS37 establishes three specific applications of these general requirements, namely:
\begin{itemize}
\item{ a provision should not be recognized for future operating losses;
}
\item{ a provision should be recognized for an onerous contract -- a contract
in which the unavoidable costs of meeting its obligations exceeds the
expected economic benefits; and
}
\item{ a provision for restructuring costs should be recognized only when an enterprise has a detailed formal plan for restructuring and has raised a valid expectation in those affected.
}
\end{itemize}

The last of these should exclude costs, such as retraining or relocating continuing staff, marketing or investment in new systems and distribution networks, the restructuring does not necessarily entail that. IAS37 requires that provisions should be recognized in the balance sheet when, and only when, an enterprise has a present obligation (legal or constructive) as a result of a past event. The event must be likely to call upon the resources of the institution to settle the obligation, and it must be possible to form a reliable estimate of the amount of the obligation. Provisions in the balance sheet should be at the best estimate of the expenditure required to settle the present obligation at the balance sheet date.  IAS37 indicates also that the amount of the provision should not be reduced by gains from the expected disposal of assets nor by expected reimbursements (arising from, for example, insurance contracts or indemnity clauses). When and if it is virtually certain that reimbursement will be received should the enterprise settle the obligation, this reimbursement should be recognized as a separate asset.

We also note the following key points relating to regulation regarding provisioning, capital and expected loss (EL) components in ``Detailed criteria 669'' \cite[p. 151]{BaselII2006}. This portion of the regulation describes a series of quantitative standards that will apply to internally generated  OpRisk  measures  for  purposes  of  calculating  the  regulatory  minimum  capital charge:
\emph{
\begin{itemize}
\item[(a)]  Any  internal  operational  risk  measurement  system  must  be  consistent  with  the  scope  of  operational  risk  defined  by  the  Committee  in  paragraph 644  and  the  loss  event types defined in Annex 9.
\item[(b)]  Supervisors  will  require  the  bank  to  calculate  its  regulatory  capital  requirement  as  the  sum  of  expected  loss  (EL)  and  unexpected  loss  (UL),  unless  the  bank  can  demonstrate that it is adequately capturing EL in its internal business practices. That is, to base the minimum regulatory capital requirement on UL alone, the bank must be  able  to  demonstrate  to  the  satisfaction  of  its  national  supervisor  that  it  has  measured and accounted for its EL exposure.
\item[(c)]  A  bank's  risk  measurement  system  must  be  sufficiently  `granular'  to  capture  the  major drivers of operational risk affecting the shape of the tail of the loss estimates.
\end{itemize}
}

Here, note that if EL was accounted for, i.e. provisioned, then it should not be covered by capital requirements again.

With regard to BEICF data, in order to understand the importance of BEICF data in OpRisk practice we discuss this data source in the form of Key Risk Indicators (KRIs), Key Performance Indicators (KPIs) and Key Control Indicators (KCIs).

A KRI is a metric of a risk factor. It provides information on the level of exposure to a given OpRisk of the organization at a particular point in time. KRIs are useful tools for business lines managers, senior management and Boards to help monitor the level of risk taking in an activity or an organization, with regard to their risk appetite.

Performance indicators, usually referred to as KPIs, measure performance or the achievement of targets. Control effectiveness indicators, usually referred to as KCIs, are metrics that provide information on the extent to which a given control is meeting its intended objectives. Failed tests on key controls are natural examples of effective KCIs.

KPIs, KRIs and KCIs overlap in many instances, especially when they signal breaches of thresholds: a poor performance often becomes a source of risk. Poor technological performance such as system downtime for instance becomes a KRI for errors and data integrity. KPIs of failed performance provide a good source of potential risk indicators. Failed KCIs are even more obvious candidates for preventive KRIs: a key control failure always constitutes a source of risk.

Indicators can be used by organizations as a means of control to track changes in their exposure to OpRisk.  When selected appropriately, indicators ought to flag any change in the likelihood or the impact of a risk occurring.
For financial institutions that calculate and hold OpRisk capital under more advanced approaches such as the previous AMA internal model approaches, KPIs, KRIs and KCIs are advisable metrics to capture BEICF. While the definition of BEICF differs from one jurisdiction to another and in many cases is specific to individual organizations, these factors must:

\begin{itemize}
\item be risk sensitive (here the notion of risk goes beyond incidents and losses);
\item provide management with information on the risk profile of the organization;
\item represent meaningful drivers of exposure which can be quantified; and
\item should be used across the entire organization.
\end{itemize}

While some organizations include the outputs of their risk and control self-assessment programs under their internal definition of BEICF's, indicators are an appropriate mechanism to satisfy these requirements, implying that there is an indirect regulatory requirement to implement and maintain an active indicator program, see discussion in \cite{Chapelle2013}.

For instance, incorporating BEICF's into OpRisk modelling is a reflection of the modelling assumption that one can see OpRisk as a function of the control environment. If the control environment is fair and under control, large operational losses are less likely to occur and OpRisk can be seen as under control. Therefore, understanding the firm's business processes, mapping the risks on these processes and assessing how the controls implemented behave is the fundamental role of the OpRisk manager. However, the SMA does not provide any real incentive mechanism firstly for undertaking such a process and secondly for incorporating this valuable information into the capital calculation.

%%%%%%%%%%%%%%%%%%%%%%%%%%%%%%%%%%%%%%%%%%%%%%%%%%%%%%%%%%%%%%%%%%%%%%%%%%%%%%
\subsection{SMA discards 75\% of OpRisk data Types}
%%%%%%%%%%%%%%%%%%%%%%%%%%%%%%%%%%%%%%%%%%%%%%%%%%%%%%%%%%%%%%%%%%%%%%%%%%%%%%
\noindent Both the Basel II and Basel III regulations emphasize the significance of incorporating a variety of loss data into OpRisk modelling and therefore ultimately into capital calculations. As has just been discussed the four primary data sources to be included are internal loss data, external loss data, scenario analysis and BEICF. However, under the new SMA framework only the first data source is utilised, the other three are now discarded.

Furthermore, even if this decision to drop BEICF's were reveresed in revisions to the SMA guidelines, we argue this would not be easy to achieve. In terms of using pieces of information such as BEICF's and scenario data, since under the SMA framework the level of model granularity is only at institution level, it does not easily lend itself to incorporation of these key OpRisk data sources.

To business lines managers, KRIs help to signal a change in the level of risk exposure associated with specific processes and activities. For quantitative modellers, key risk indicators are a way of including BEICFs into OpRisk capital. However, since BEICF data does not form a component of required data for the SMA model, there is no longer a regulatory requirement or incentive under the proposed SMA framework to make efforts to develop such BEICF data sources. This not only reduces the effectiveness of the risk models through the loss of a key source of information, but in addition the utility of such data for risk management practitioners and managers is reduced as this data is no longer collected with the same required scrutiny, including validation, data integrity and maintenance, and reporting that was previously required for AMA internal models using such data.

These key sources of OpRisk data are not included in the SMA and  cannot easily be incorporated into an SMA framework even if there were a desire to do so due to the level of granularity implied by the SMA. This makes capital calculations less risk sensitive. Furthermore, the lack of scenario based data incorporated into the SMA model makes it less forward looking and anticipatory as an internal model based capital calculation framework.

%%%%%%%%%%%%%%%%%%%%%%%%%%%%%%%%%%%%%%%%%%%%%%%%%%%%%%%%%%%%%%%%%%%%
%%%%%%%%%%%%%%%%%%%%%%%%%%%%%%%%%%%%%%%%%%%%%%%%%%%%%%%%%%%%%%%%%%%%
\section{SMA can be a super-additive capital calculation }\label{superadditvity1_sec}
The SMA seems to have the unfortunate feature that it may produce capital at a group level compared to the institutional level in a range of jurisdictions which has the property that it is super-additive. It might be introduced by the regulator on purpose to encourage splitting of very large institutions though it is not stated in the Basel Committee documents explicitly. In this section we show several examples of super-additivity and discuss its implications.

\subsection{Examples of SMA super-additivity}
\noindent Consider two banks with identical $BI$ and $LC$. However, the first bank has only one entity where as the second has two entities. The two entities of the second bank have the same $BI$ and the same $LC$, and those are equal to both half the $BI$ and half the $LC$ of the first joint bank.

\begin{figure}[!h]\begin{center}
\captionsetup{width=0.95\textwidth}
\includegraphics[scale=0.8]{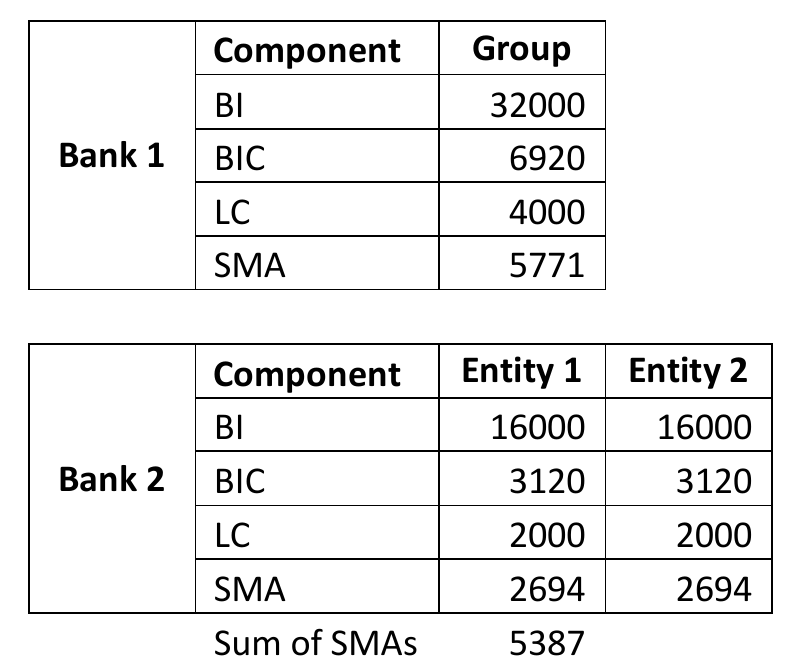}\hspace{0.6cm}\includegraphics[scale=0.8]{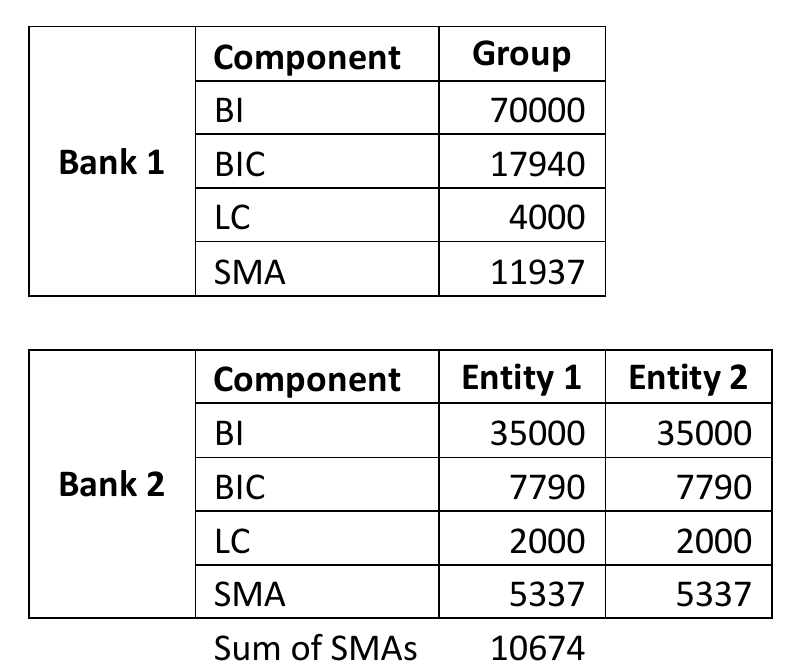}
\caption{\footnotesize{{Super-additivity examples (all amounts are in Euro million). Left figure -- the case of bucket shift: Bank 1 is in Bucket 5 and the entities of Bank 2 are in Bucket 4. Right figure -- the case of no bucket shift:  Bank 1 and the entities of Bank 2 are in Bucket 5.}}} \label{Fig_SubadditvityExamples1and2}
\end{center}
\end{figure}

In case one, Figure \ref{Fig_SubadditvityExamples1and2} (left side), we consider the situation of a bucket shift, where the SMA capital obtained for the joint bank is Euro 5,771 million while the sum of the SMA capital obtained for the two entities of the second bank is only Euro 5,387 million. In this example, the SMA does not capture a diversification benefit, on the contrary, it assumes that the global impact of an incident is larger than the sum of the parts. Here, the joint bank is in Bucket 5 while the entities appear in Bucket 4.
In the second case, Figure \ref{Fig_SubadditvityExamples1and2} (right side), we consider no bucket shift between the joint bank (Bank 1) and the two entity bank (Bank 2). Bank 1 is in Bucket 5 and the entities of Bank 2 are in Bucket 5 too. In this case we see that the joint bank has an SMA capital of Euro 11,937 million, whereas the two entity bank has an SMA capital of Euro 10,674 million. Again there is a super-additive property.

Of course in the above examples we set $BI$ and $LC$ somewhat arbitrary. So, in the next example we use $BI$ implied by the 0.999 VaR of LDA. In particular, assume a bank with $Poisson(\lambda)-Lognormal(\mu,\sigma)$ risk profile at the top level. Then calculate the long term average $LC$ using (\ref{LTALC_LN_eq}) and the 0.999 VaR of the annual loss using (\ref{SLALN}), and find the implied $BI$ by matching SMA capital with the 0.999 VaR. Now, consider the identical Bank 2 that splits into two similar independent entities that will have the same $LC$ and the same $BI$ both equal to half of $LC$ and half of $BI$ of the 1st bank that allows to calculate SMA capital for each entity. Also note that in this case, the entities will have risk profiles $Poisson(\frac{1}{2}\lambda)-Lognormal(\mu,\sigma)$ each.

 \begin{remark}\label{sum_compoundPorcess_remark} The sum of $K$ independent compound  processes $Poisson(\lambda_i)$ with severity $F_i(x)$, $i=1,\ldots,K$ is a compound process $Poisson(\lambda)$ with $\lambda=\lambda_1+\cdots+\lambda_K$ and severity $F(x)=\frac{\lambda_1}{\lambda}F_1(x)+\cdots+\frac{\lambda_K}{\lambda}F_K(x)$; see e.g. \cite[section 7.2, proposition 7.1]{Shevchenko2011}.
 \end{remark}

\begin{figure}[!h]\begin{center}
\captionsetup{width=0.95\textwidth}
\includegraphics[scale=0.8]{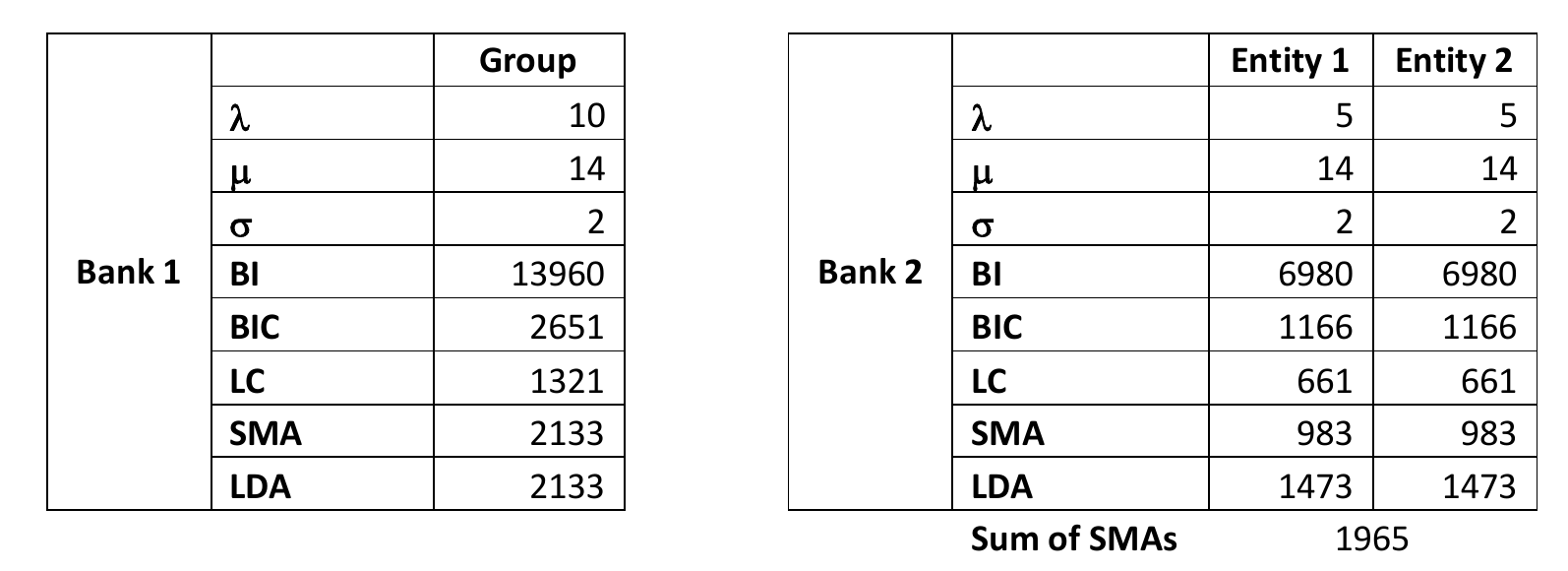}
\caption{\footnotesize{{Super-additivity example (all amounts are in Euro million). $BI$ for Bank 1 is implied by the 0.999 VaR of $Poisson(\lambda)-Lognormal(\mu,\sigma)$ risk profile (LDA).}}} \label{Fig_SubadditvityExamples3}
\end{center}
\end{figure}

 The results in the case of $\lambda=10$, $\mu=14$, $\sigma=2$ are shown in Figure \ref{Fig_SubadditvityExamples3}.
Here, SMA for Bank 1 is Euro 2.13 billion while the sum of SMAs of the entities of Bank 2 is Euro 1.96 billion demonstrating the sub-additivity feature. Note that in this case one can also calculate the 0.999 VaR for each entity, which is Euro 1.47 billion, while SMAs for Entity 1 and Entity 2 are Euro 983 million each. That is, the entities with SMA capital are significantly undercapitalized when compared to the LDA economic capital model; this subject will be discussed more in Section \ref{superadditvity2_sec}.

%\begin{table}[!h]\begin{center}
%\captionsetup{width=0.7\textwidth}
%\includegraphics[scale=0.8]{Table6}
%\caption{{{Super-additivity issue illustrated (in million). Bank 1 and the entities of Bank 2 are in Bucket 5.}}} \label{tab6}
%\end{center}
%\end{table}

Next we state a mathematical expression that a bank could utilize in business structure planning to decide, in the long term, if it will be advantageous under the new SMA framework to split into two entities (or more) or remain in a given joint structure, according to the cost of funding Tier I SMA capital.

Consider the long term SMA capital behavior averaged over the long term history of the SMA capital for each case, joint and disaggregated business models. Then, from the perspective of a long term analysis regarding restructuring, the following expressions can be used to determine the point at which the SMA capital would be super-additive. If it is super-additive in the long term, it would indicated that there is therefore an advantage to split the institution in the long-run into disaggregated separate components. Furthermore, the expression provided allows one to maximize the long term SMA capital reduction that can be obtained under such a partitioning of the institution into $m$ separate disaggregated entities.

\begin{proposition}[Conditions for super-additive SMA Capital] \label{propSuperAdd}
Under the LDA models with the frequency from Poisson($\lambda_J$) and generic severity $F_X(x;\bm{\theta}_J)$, the long term average of the loss component $\widetilde{LC}$, can be found using,
\begin{equation}
\widetilde{LC}=\lambda\times\bigg(7\, \mathrm{E}[X]+7\, \mathrm{E}[X|X>L]+5\, \mathrm{E}[X|X>H]\bigg),
\end{equation}
i.e. the short-term empirical loss averages in SMA formula (\ref{LC_eq}) are  replaced with the `long-term' average. We now denote the long term average LC for a bank as $\widetilde{LC}(\lambda_J,\bm{\theta}_J)$ and for the $m$ bank entities after split as $\widetilde{LC}(\lambda_i,\mu_i,\sigma_i)$, $i\in\left\{1,2,\ldots,m\right\}$. Therefore the long term SMA capital $K_{SMA}(BI_J,\widetilde{LC}(\lambda_J,\bm{\theta}_J))$ will be an explicit function of LDA model parameters $(\lambda_J,\bm{\theta}_J)$  and the long term SMA capital for the entity $i$ is  $K_{SMA}(BI_i,\widetilde{LC}(\lambda_i,\bm{\theta}_i))$. Hence, the SMA super-additive capital condition becomes:
\begin{equation}
K_{SMA}(BI_J,\widetilde{LC}(\lambda_J,\bm{\theta}_J)) - \sum_{i=1}^m K_{SMA}(BI_i,\widetilde{LC}(\lambda_i,\bm{\theta}_i)) > 0.
\end{equation}
\end{proposition}

 The above condition is for a model based restructuring assuming each bank entity is modelled generically by a LDA model. Structuring around such a model based assumption can be performed to determine optimal disaggregation of the institution to maximize capital cost reductions. Many severity distribution types will allow calculation of the long-term $LC$ in closed form. For example, in the case of Poisson-Lognormal model it is given by (\ref{LTALC_LN_eq}).

Of course, one can also use the above condition to perform  maximization of the capital reduction over the next year by replacing $\widetilde{LC}$ with the observed $LC$ calculated from empirical sample averages of historical data as required in the SMA formula avoiding explicit assumptions for severity and frequency distributions distributions.
%This condition can have many possible uses, for instance if the bank has historical data for 5-10 years (or more) as required by the SMA, then they can substitute for the long term averages, the empirical sample averages. This will allow them to identify an optimal split to maximise the capital cost reduction such that the positive difference between $K_{SMA}(BI_J,\widetilde{LC}(\lambda_J,\bm{\theta}_J))$ and $\sum_{i=1}^m K_{SMA}(BI_i,\widetilde{LC}(\lambda_i,\bm{\theta}_i))$ is greatest.

%If instead one had chosen a Poisson-Pareto model, with Pareto distribution characterized by survival function $\overline{F}_X(x;\alpha, x_0)$, given by
%\begin{equation}
%\overline{F}_X(x;\alpha,x_0) = \begin{cases}
%\left(\frac{x_0}{x}\right)^{\alpha}, & x \geq x_0,\\
%1, & x \leq x_0,
%\end{cases}
%\end{equation}
%then the long-term Loss Component is well defined when $\alpha > 1$ and given by
%\begin{equation}
%\widetilde{LC}(\lambda,\alpha, x_0) = \lambda\times \left( \frac{\alpha (7 x_0 + 7 L + 5 H) }{\alpha - 1}\right).
%\end{equation}

%%%%%%%%%%%%%%%%%%%%%%%%%%%%%%%%%%%%%%%%%%%%%%%%%%%%%%%%%%%%%%%%%%%%
%%%%%%%%%%%%%%%%%%%%%%%%%%%%%%%%%%%%%%%%%%%%%%%%%%%%%%%%%%%%%%%%%%%%
\subsection{SMA super-additivity, macro-prudential policy \& systemic risk}\label{superadditvity2_sec}
%%%%%%%%%%%%%%%%%%%%%%%%%%%%%%%%%%%%%%%%%%%%%%%%%%%%%%%%%%%%%%%%%%%%
%%%%%%%%%%%%%%%%%%%%%%%%%%%%%%%%%%%%%%%%%%%%%%%%%%%%%%%%%%%%%%%%%%%%
\noindent In this section we discuss the fact that the financial system is not invariant under observation, that is banks and financial institutions will respond in a rational manner to maximize their competitive advantage. In particular, if new regulations allow and indeed provide incentives for banks to take opportunities to reduce for instance the cost of capital they will generally act to do so. It is in this context that we introduce in brief the relationship between the new SMA capital calculations and the broader macro-prudential view of the economy that the regulator holds.

It is generally acknowledged that the enhancement of the Basel II banking regulations by the additions that Basel III accords brought to the table, were largely driven by a desire to impose greater macro-prudential oversight on the banking system post the 2008 financial crisis. Indeed, the Basel III accords adopted an approach to financial regulation aimed at mitigating the ``systemic risk'' of the financial sector as a whole, where we may adopt a generic high level definition of systemic risk as
\begin{quotation}
``\textsl{...the disruption to the flow of financial services that is (i) caused by an impairment of all or parts of the financial system; and (ii) has the potential to have serious negative consequences for the real economy....}''
\end{quotation}
This view of systemic risk is focused on disruptions that arise from events such as the collapse of core banks or financial institutions in the banking sector, such as what happened post the Lehman Brothers collapse leading to a wider systemic risk problem of the 2008 financial crisis.

In response to reducing the likelihood of such a systemic risk occurrence, the Basel III regulation imposed several components that are of relevance to macro-prudential financial regulation. Under Basel III, banks' capital requirements have been strengthened and new liquidity requirements, a leverage cap and a countercyclical capital buffer were introduced; these would remain in place under the new SMA guidelines.

In addition, large financial institutions, i.e. the largest and most globally active banks, were required to hold a greater amount of capital with an increase proportion of this Tier I capital being more liquid and of greater credit worthiness, i.e. `higher-quality' capital. We note that this is consistent with a view of systemic risk reduction based on a cross-sectional approach. For instance, under this approach the Basel III requirements sought to introduce systemic risk reduction macro-prudential tools such as: a) countercyclical capital requirements which were introduced with the purpose to avoid excessive balance-sheet shrinkage from banks in distress that may transition from going concerns to gone concerns; b) caps on leverage in order to reduce or limit asset growth through a mechanism that linked a banks' assets to their equity; c) time variation in reserve requirements with pro-cyclical capital buffers as a means to control capital flows with prudential purposes. In the UK, such Basel III macro-prudential extensions are discussed in detail in the summary of the speech given at the 27th Annual Institute of International Banking Conference, Washington, by the Executive Director for Financial Stability Strategy and Risk, in the Financial Policy Committee at the Bank of England \footnote{\url{http://www.bankofengland.co.uk/publications/Documents/speeches/2016/speech887.pdf}}.

Furthermore, one can argue that several factors can contribute to the systemic risk build up in an economy, locally and globally. One of these of relevance to discussions on AMA internal modelling versus SMA models is the risk that arises from complexity of mathematical modelling being adopted in risk management and product structuring/pricing. One can argue from a statistical perspective that to scientifically understand the complex nature of OpRisk processes and then to respond to them with adequate capital measures and risk mitigation policies and governance structuring it would be prudent to invest in some level of model complexity. However, with such complexity comes a significant chance of misuse of such models for gaming of the system to obtain competitive advantage via for instance achieving a reduction in capital. This can inherently act as an unseen trigger for systemic risk, especially if it is typically taking place in the larger, more substantial banks in the financial network as is the case under AMA Basel II internal modelling. Therefore, we have this tension between reducing the systemic risk in the system due to model complexity and actually understanding scientifically the risk processes. We argue that the SMA goes too far in simplifying the complexity of OpRisk modelling, rendering it unusable for risk analysis and interpretation. However, perhaps model complexity reduction could instead be reduced through AMA standardization of internal modelling practice, something we will discuss in the conclusions of this paper.

In this section we ask what role can the SMA model framework play in the context of macro-prudential systemic risk reduction. To address this question, we adopt the SMA's highly stylised view as follows. We first consider the largest banks and financial institutions in the world. These entitites are global, and key nodes in the financial network, sometimes they have been referred to as the so called ``too big to fail'' institutions. It is clear that the existence of such large financial institutions has both positive and negative economic effects. However, from a systemic risk perspective, they can pose problems for banking regulations both in local jurisdictions and globally.

There is in general an incentive to reduce the number of such dominant nodes in the banking financial network when viewed from the perspective of reducing systemic risk. So the natural question that arises with regard to the SMA formulation: \emph{does this new regulation incentivise disaggregation of large financial institutions and banks, at least from a high level perspective of reduction of the costs associated with obtaining, funding and maintaining Tier I capital and liquidity ratios required under Basel III at present?} In addition, if a super-additive capital is possible: is it achievable for feasible and practically sensible disaggregated entities? Thirdly, one could ask: does this super-additive SMA capital feature provide an increasing reward in terms of capital reduction as the size of the institution increases? We address these questions in the following stylized case studies that can illustrate that in fact the SMA can be considered as a framework that will induce systemic risk reductions from the perspective of providing potential incentives to reduce capital costs through disaggregation of large financial institutions and banks in the global banking network. \emph{However, we also observed that this may lead to significant under-capitalization of the entities after disaggregation; for example, in the case already discussed in Figure \ref{Fig_SubadditvityExamples3} and considered in the next section Figure \ref{FigSuperadditvityUndercapitalization}}.

\subsection{SMA super-additivity is feasible and produces viable BI}
%%%%%%%%%%%%%%%%%%%%%%%%%%%%%%%%%%%%%%%%%%%%%%%%%%%%%%%%%%%%%%%%%%%%
\noindent For illustration, assume the joint institution is simply modelled by a Poisson-Lognormal model Poisson($\lambda_J$)-Lognormal($\mu_J,\sigma_J$) with parameters sub-indexed by $J$ for the joint institution and a $BI$ for the joint institution denoted by $BI_J$. Furthermore, we assume that if the institution had split into $m=2$ separate entities for Tier I capital reporting purposes then each would have its own stylized annual loss modelled by two independent Poisson-Lognormal models: Entity 1 modelled by Poisson($\lambda_1$)-Lognormal($\mu_1,\sigma_1$) and Entity 2 modelled by Poisson($\lambda_2$)-Lognormal($\mu_2,\sigma_2$) with $BI_1$ and $BI_2$, respectively. Here we assume that the disaggregation of the joint institution can occur in such a manner that the risk profile of each individual entity may adopt more, less or equal risk aversion, governance and risk management practices. This means that there is really no restrictions of the parameters $\lambda_1$, $\mu_1$ and $\sigma_1$ nor on parameters $\lambda_2$, $\mu_2$ and $\sigma_2$ from the perspective of $\lambda_J$, $\mu_J$ and $\sigma_J$.

In this sense we study the range of parameters and $BI$ values that will provide incentive for large institutions to reduce systemic risk by undergoing disaggregation into smaller institutions. We achieve this through consideration of the SMA super-additivity condition in Proposition \ref{propSuperAdd}. In this case it leads us to consider
\begin{equation}\label{superadditivity_cond2}
K_{SMA}(BI_J,\widetilde{LC}(\lambda_J,\mu_J,\sigma_J))-\sum _{i=1}^2 K_{SMA}(BI_i,\widetilde{LC}(\lambda_i,\mu_i,\sigma_i))>0.
\end{equation}

Using this stylized condition, banks may be able to determine for instance if in the long term it would be economically efficient to split their institution into two or more separate entities. Furthermore, they can use this expression to optimize the capital reduction for each of the individual entities, relative to the combined entities SMA capital. Hence, what we show here is the long term average behavior which will be the long run optimal conditions for split or merge.

We perform a simple analysis below where at present the joint institution is modelled by an LDA model with frequency given by $Poisson(\lambda_J=10)$ and severity given by $Lognormal(\mu_J = 12, \sigma_J = 2.5)$, and with a $BI$ implied by the 0.999 VaR of LDA model, as detailed in Table \ref{ImpliedBI_tab}, giving $BI = \mbox{Euro }14.24$ billion. We then assume the average number of losses in each institution if the bank splits into two is given by $\lambda_1 = 10$ and $\lambda_2 = 10$, also the scale of the losses changes but the tail severity of large losses is unchanged such that $\sigma_1 = 2.5$ and $\sigma_2 = 2.5$ but $\mu_1$ and $\mu_2$ are unknown. In addition, we calculate $BI_1$ and $BI_2$ implied by the LDA 0.999 VaR for the given values of $\mu_1$ and $\mu_2$.
%We assume that the joint institution is sufficiently large that it is in Bucket 2 or above and that any split in the institution into two independent entities is performed only if the entities remain of sufficient volume to be in Bucket 2 or above after disaggregation.
Then we determine the set of values of $\mu_1$ and $\mu_2$ such that condition (\ref{superadditivity_cond2}) is satisfied.

%To achieve this we observe that the joint SMA capital is fixed and given by the BI value such that we have the following closed form expression for the SMA capital when matched to the LDA VaR at quantile level $\alpha$ according to Equation \ref{SLALN} to give:
%\begin{equation} \label{SMASLAmatch}
%SMA(\lambda_J,\mu_J,\sigma_J,BI_J) = SLA(\alpha;\lambda,\mu,\sigma),
%\end{equation}
%where we consider in these examples $\alpha = 0.999$.
%
%In the case of disaggregation of the institution into two components, we undertake an analysis in which we explore the values of $\mu_1$ and $\mu_2$ such that we undertake the following steps to determine if the combination of $\mu_1$ and $\mu_2$ will satisfy the inequality in Proposition \ref{propSuperAdd}:
%\begin{enumerate}
%%
%\item Construct a set of tuples of $\mu_1$ and $\mu_2$ given by $\left\{(\mu_1,\mu_2): \mu_1 \in [7,15] \, \& \, \mu_2 \in [7,15].\right\}$ based on a grid.
%%
%\item Evaluate for each grid point combination of $(\mu_1,\mu_2)$ the value of $BIC_1$ and $BIC_2$ that will ensure that equation \ref{SMASLAmatch} holds for $SMA(\lambda_1,\mu_1,\sigma_1,BI_1) = SLA(\alpha;\lambda_1,\mu_1,\sigma_1)$
%and \\$SMA(\lambda_2,\mu_2,\sigma_2,BI_2) = SLA(\alpha;\lambda_2,\mu_2,\sigma_2)$
%%
%\item Given the $(\mu_1,\mu_2)$ and obtained $(BIC_1,BIC_2)$ combinations, determine if
%\begin{equation}
%K_{SMA}(BI_J,\widetilde{LC}(\lambda_J,\mu_J,\sigma_J)) -K_{SMA}(BI_1,\widetilde{LC}(\lambda_1,\mu_1,\sigma_1)) -K_{SMA}(BI_2,\widetilde{LC}(\lambda_2,\mu_2,\sigma_2)) >0,
%\end{equation}
%is satisfied or not for this combination.
%%
%\end{enumerate}

 Table \ref{tab_superadditvityBI} shows the range of values for which  $\mu_1$ and $\mu_2$  will produce super-additive capital structures and therefore justify disaggregation from the perspective of SMA capital cost minimization. We learn from this analysis that indeed it is plausible to structure a disaggregation of a large financial institution into a smaller set of financial institutions under the SMA capital structure. That is the range of parameters $\mu_1$ and $\mu_2$ that produce super-additive capital structures under the SMA formulation set-up are plausible, and the $BI$ values are plausible for such a decomposition.
In Table \ref{tab_superadditvityBI} we also show the $BI$ values for the Entity 1 implied in the cases satisfying the super-additive SMA capital condition.

%\begin{center}
%\begin{figure}[!h]
%\captionsetup{width=0.9\textwidth}
%\begin{center} \includegraphics[scale=0.6]{Figure_1_SupperAdd.pdf}%[width = 10cm, height = 10cm]
%\end{center}
%\caption{\footnotesize{Range of model parameters related to disaggregated bank entities ($\mu_1$ and $\mu_2$) that produce super-additive capital in white (black area corresponds to no super-additvity).}}\label{SuperadditiveCases_fig}
%\end{figure}
%\end{center}

\begin{table}[!h]\begin{center}
\captionsetup{width=0.9\textwidth}
\caption{\footnotesize{{Implied $BI$ in billion of Euro for Entity 1 (N.A. indicates no super-additive solution).}}} \label{tab_superadditvityBI}
{\footnotesize{
\begin{tabular*}{0.52\textwidth}{c|cccccc} \toprule
$\mu_1\setminus\mu_2$ & 8 & 9 & 10 & 11 & 12&  13 \\
 \midrule
8  &   0.301 &  0.301  &  0.301 &   0.301 &   N.A. &  N.A.  \\
9  &   0.820 &   0.820  &  0.820 &   0.820 &  N.A.  &       N.A.\\
10 &  2.406 &   2.406  &  2.406 &  2.406 &   N.A.  &       N.A.\\
11 &   5.970 &   5.970  &  5.970 &  5.970 &   N.A.  &       N.A.\\
12 &   N.A. &   N.A.  &  N.A. &     N.A. &     N.A.  &       N.A.\\
13 &     N.A. &     N.A.  &    N.A. &     N.A. &     N.A.  &       N.A.\\
\bottomrule
\end{tabular*}
}}
\end{center}
\end{table}

We see from these results that the range of $BI$ values that are inferred under the super-additive capital structures are also plausible ranges of values. This demonstrates that, it is practically feasible for the SMA to produce incentive to reduce capital by disaggregation of larger institutions into smaller ones.

%%%%%%%%%%%%%%%%%%%%%%%%%%%%%%%%%%%%%%%%%%%%%%%%%%%%%%%%%%%%%%%%%%%%
\subsection{Does SMA incentivise Larger Institutions to Disaggregate more than Smaller Institutions?}
%%%%%%%%%%%%%%%%%%%%%%%%%%%%%%%%%%%%%%%%%%%%%%%%%%%%%%%%%%%%%%%%%%%%
\noindent Finally, we complete this section by addressing the question:
\textsl{Is there an increasing in potential capital reductions through using super-additive SMA capital as a motivation to disaggregate into smaller firms as the size of the institution increases?}
As the tail of the bank loss distribution increases, the size of the institution as quantified through $BI$ increases and we are interested to see if there is increasing incentive for larger banks to disaggregate to reduce capital charge.

%This leads us to consider a class of LDA model in which the VaR (capital risk measure) is known to be super additive and furthermore the loss model is stochastically ordered in its tail index. The reason for this is that

To illustrate this point we perform a study in which we take the following set-up. We assume we have a bank with $Poisson(\lambda)-Lognormal(\mu,\sigma)$ operational risk profile at the institution level. Now we calculate $\widetilde{LC}$ and match LDA VaR capital measure via the single loss approximation at 99.9\% quantile to the SMA capital to imply the corresponding $BI$, $BIC$ and $SMA=K_{SMA}(BI,\widetilde{LC})$ capital. Next we consider to disaggregate the institution into two similar independent components that we denote as Entity 1 and Entity 2. This means the the entities will have $\widetilde{LC}_1=\widetilde{LC}_2=\frac{1}{2}\widetilde{LC}$, $BI_1=BI_2=\frac{1}{2}BI$, $\lambda_1 = \lambda_2 = \frac{1}{2}\lambda$, $\mu_1=\mu_2=\mu$ and $\sigma_1=\sigma_2=\sigma$; see Remark \ref{sum_compoundPorcess_remark}. Then we calculate the SMA capitals $SMA_1=K_{SMA}(BI_1,\widetilde{LC}_1)$ and $SMA_2=K_{SMA}(BI_2,\widetilde{LC}_2)$  for Entity 1 and Entity 2, and find the absolute super-additivity benefit from SMA capital reduction $\Delta =SMA-SMA_1-SMA_2$ and relative benefit $\Delta/SMA$.

\begin{figure}[!h]
\captionsetup{width=0.9\textwidth}
\begin{center}\includegraphics[scale=0.65]{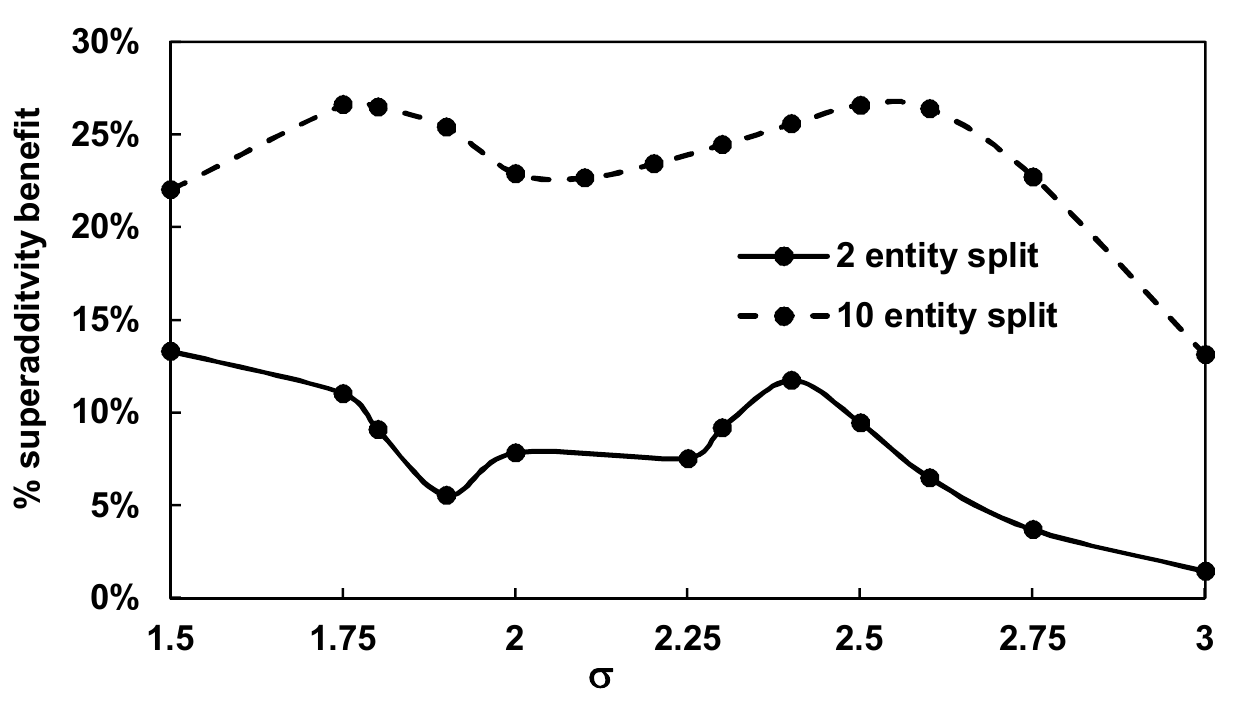}\includegraphics[scale=0.65]{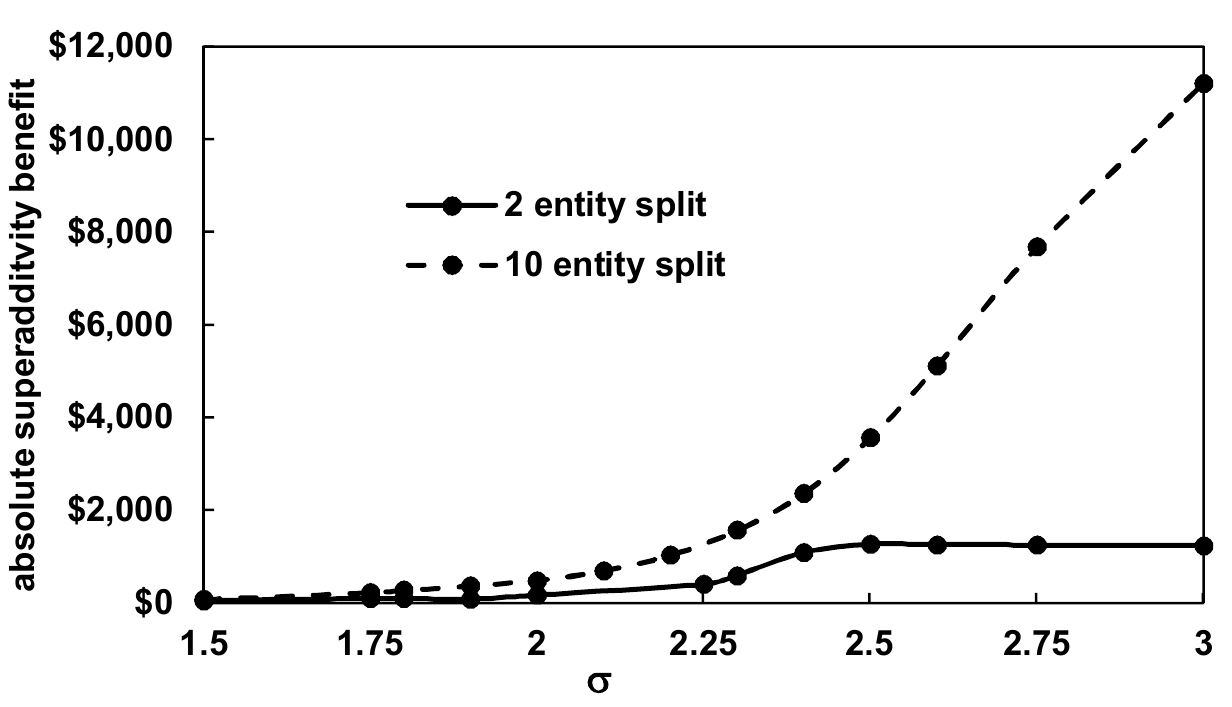}
\caption{\footnotesize{Absolute (in Euro million) and relative super-additivity benefits from splitting a bank into two similar entities (and into ten similar entities) versus Lognormal severity shape parameter $\sigma$.}}\label{FigSuperadditvityBenefit}
\end{center}
\end{figure}

 The results in the case $\lambda=10$, $\mu=14$ and varying $\sigma$ are plotted in Figure \ref{FigSuperadditvityBenefit}.
Note that the benefit is non-monotonic function of $\sigma$, because in some cases the disaggragation process results in the entities shifting into a different SMA capital bucket compared to the original joint entity.
One can also see that the absolute benefit from a bank disaggregation increases as the bank size increases though the relative benefit drops. On the same figure we also show the results for the case of bank disaggregation into 10 similar independent entities, i.e. $\lambda_1 =\cdots= \lambda_{10}=\lambda/10$, $\mu_1=\cdots=\mu_{10}=\mu$ and $\sigma_1=\cdots=\sigma_{10}=\sigma$.

\begin{figure}[!h]
\captionsetup{width=0.9\textwidth}
\begin{center}\includegraphics[scale=0.65]{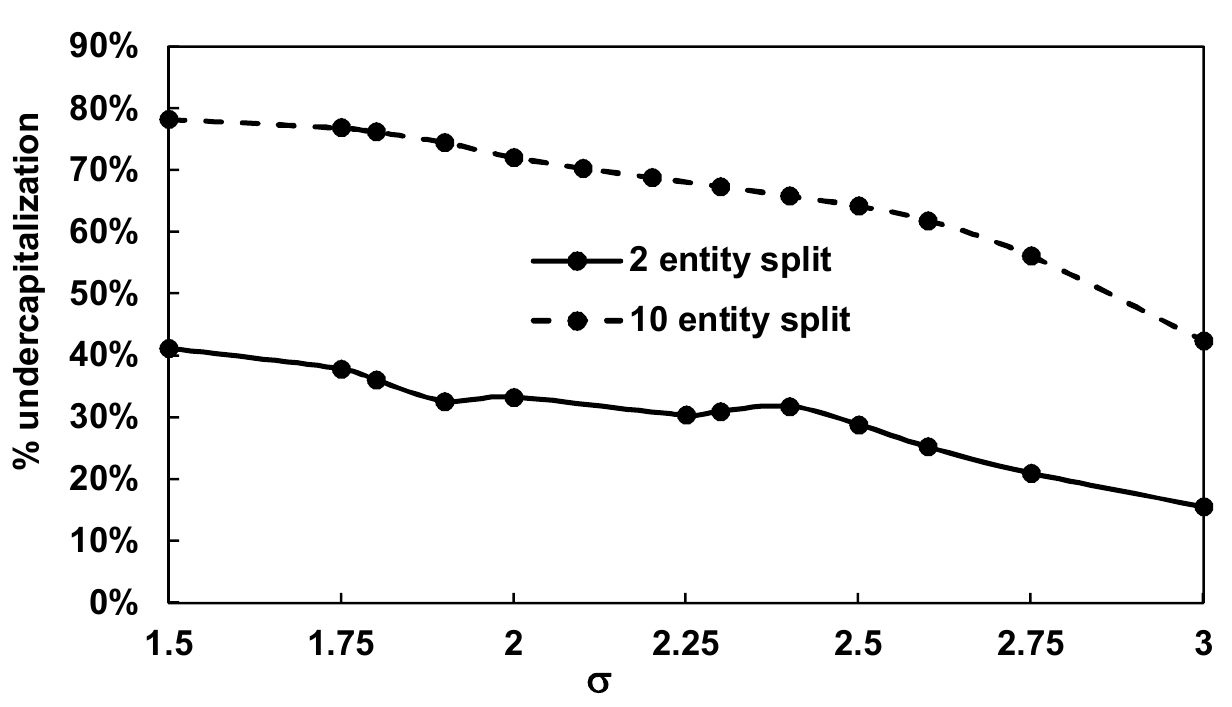}\includegraphics[scale=0.65]{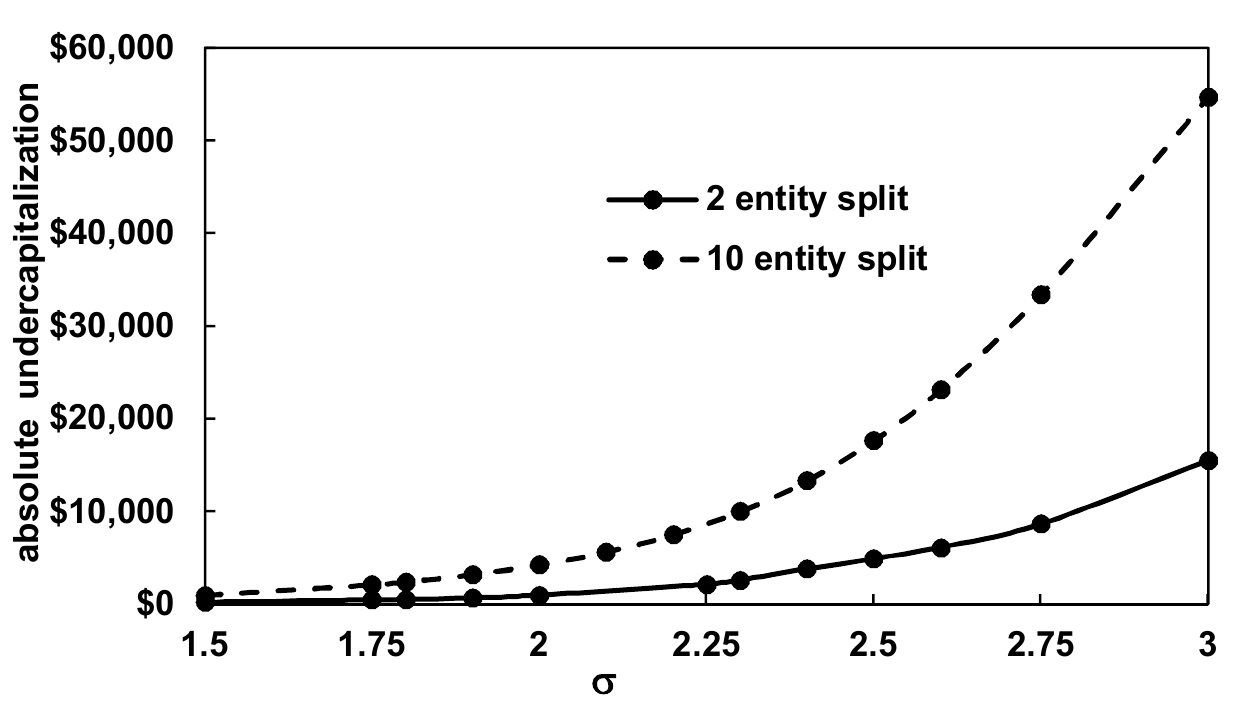}
\caption{\footnotesize{Absolute (in Euro million) and relative under-capitalization of the entities after bank disaggregation into two similar entities (and into ten similar entities) versus Lognormal severity shape parameter $\sigma$.}}\label{FigSuperadditvityUndercapitalization}
\end{center}
\end{figure}

We also calculate the 0.999 VaR of the Poisson-Lognormal process denoted as $LDA_1$ and $LDA_2$ for Entity 1 and Entity 2 respectively using single loss approximation (\ref{SLALN}). Then we find under-capitalization of the entities $LDA_1+LDA_2-SMA_1-SMA_1$ introduced by disaggregation and corresponding relative under-capitalisation $(LDA_1+LDA_2-SMA_1-SMA_1)/(LDA_1+LDA_2)$. These results (and also for the case of bank disaggregation into 10 similar entities) are shown in Figure \ref{FigSuperadditvityUndercapitalization}.
In this example, under-capitalization is very significant and increases for larger banks though the relative under-capitalization is getting smaller.
Moreover, both the super-additivity benefit and under-capitalization features become more pronounced in the case of splitting into 10 entities when compared to the two-entity split.

Thus, from one perspective the capital calculated using SMA formula encourages large banks to disaggregate (reducing systemic risk from failure in the banking network) but from another side introduces significant under-capitalization of newly formed smaller entities increasing their chances of failure (i.e. increasing systemic risk in the banking system). In light of our analysis it becomes clear that there is a downward pressure on banks to disaggregate, hence reducing some aspects of systemic risk in the banking network. However, we also show that at the same time, this very mechanism may under-capitalise the resulting smaller institutions which would in turn increase systemic risk. The final outcome of the stability of the banking network will therefore depend largely on how aggressively larger banks choose to seek capital reductions at the risk of under-capitalising their disaggregated entities, i.e. their risk appetite in this regard will dictate the ultimate outcome. Surely, such an uncertain future is not what the regulator would have had in mind in allowing for an incentive for disaggregation through the existence of super-additive capital measures.

%To conclude we can also say mathematically some nice properties of super-additivity versus bank size that complement the study undertaken above. A good example of such a model of relevance to this case study is the Poisson-Pareto model. For such a model we can imagine the joint institution as being represented by a Pareto severity model which can be represented according to the following regular variation asymptotic right tail representation
%\begin{equation}
%F(x) = x^{-\alpha}L(x)[1 + o(1)], \, x > 0, \, \alpha > 0
%\end{equation}
%where $L$ is a slowly varying function (eg. a logarithm). Now if we consider the joint financial institution to be a Poisson-Pareto model with severity model with tail index $\alpha$. In addition, consider the individual institutions that the joint institution would split into to be also modelled by severity models with regularly varying tails, say also Pareto models with non-degenerate tail indexes $\alpha$. In the case that $\alpha < 1$ one can find that the capital of the institution will be super-additive leading to incentive to disaggregate. Furthermore, as $\alpha$ decreases away from 1 towards 0 the size of the bank increases according to the BI implied by the SMA. Furthermore, the gain in capital saving from disaggregation also increases. This implies that the SMA gives increasing incentive to larger banks that smaller banks who are in the super-additive capital structure cases to disaggregate.

%%%%%%%%%%%%%%%%%%%%%%%%%%%%%%%%%%%%%%%%%%%%%%%%%%%%%%%%%%%%%%%%%%%%
\section{OpCar Estimation Framework}\label{OpCarEstimation_sec}
%%%%%%%%%%%%%%%%%%%%%%%%%%%%%%%%%%%%%%%%%%%%%%%%%%%%%%%%%%%%%%%%%%%%
%%%%%%%%%%%%%%%%%%%%%%%%%%%%%%%%%%%%%%%%%%%%%%%%%%%%%%%%%%%%%%%%%%%%
 This section summarizes details of the OpCar model which is the precursor to the SMA model and helped to form the SMA structure. We point out firstly that the proposed OpCar model \cite[annex 2]{BCBSd2912014} is based on LDA, albeit a very simplistic one that models the annual loss of the institution as a single LDA model formulation
%This is clear where the OpCar is described as ``\textsl{...The OpCaR methodology estimates a bank's OpRisk capital through the convolution of a single severity distribution and a single frequency distribution. Each bank's OpCaR estimate was assumed to refer to a unique OpRisk category, having a specific aggregated frequency and severity of losses...}''
\begin{equation}
Z=\sum_{i=1}^N X_i,
\end{equation}
where $N$ is the annual number of losses modelled as random variable from Poisson distribution, $Poisson(\lambda)$, i.e. $\lambda=E[N]$, and $X_i$ is loss severity random variable from distribution $F_X(x;\bm\theta)$ parameterised by vector  $\bm\theta$. It is assumed that $N$ and $X_i$ are independent and $X_1,X_2,\ldots $ are independent too (note that modelling severities with autocorrelation is possible, see e.g. \cite{guegan2013using}. $F_X(x;\bm\theta)$ is modelled by one of the following \emph{two-parameter} distributions: \emph{Pareto, Lognormal, Log-logistic, Log-gamma}. Three variants of the Pareto model were considered as informally described in the regulation as corresponding to Pareto-light,  Pareto-medium and Pareto-heavy. As a result, up to six estimates of the 0.999 VaR were generated per bank that were averaged to find the final capital estimate used in the regression model.
%We note that three variants of the Pareto model were considered as informally described in the regulation as corresponding to relatively ``light, intermediate and heavy tailed''. In this paper it will be sufficient to present results based on the example for the Poisson-Lognormal model. Historically, this has been the most widely utilised model in OpRisk modelling practice. We consider the following parameterization:
%\begin{itemize}
%%
%\item Lognormal density:
%\begin{equation}
%f_X(x; \mu,\sigma) = \frac{1}{x\sigma\sqrt{2\pi}}\exp\left(-\frac{(\ln x - \mu)^2}{2\sigma^2}\right), \, x > 0, \, \mu \in (-\infty,\infty),\, \sigma \geq 0;
%\end{equation}
%%
%\end{itemize}
Next, we outline the OpCar fitting procedure and highlight potential issues.
%For completeness to see how the OpCar model is calibrated and to highlight some potential issues with this framework which led to the SMA model we describe below the fitting procedure for the LDA models that were fit for each bank according to the QIS LDCE results for the six OpCar models.

%%%%%%%%%%%%%%%%%%%%%%%%%%%%%%%%%%%%%%%%%%%%%%%%%%%%%%%%%%%%%%%%%%%%%%%%%%%%%%
\subsection{Parameter Estimation}\label{ParamEstimationOpCar_sec}
%%%%%%%%%%%%%%%%%%%%%%%%%%%%%%%%%%%%%%%%%%%%%%%%%%%%%%%%%%%%%%%%%%%%%%%%%%%%%%
%Now that we have seen where the approximation in the OpCar framework comes from and how its accuracy behaves asymptotically, we can continue to see how it was used in developing the new revised OpRisk modelling guidelines. In this regard, we note that the proposed OpCar model is fit using the following data for each bank in the data set over $T=5$ years:
The proposed OpCar model is estimated using data collected in Quantitative Impact Study (QIS) performed by the Basel Committee.  Specifically, for each bank in the dataset over $T=5$ years corresponding to the 2005--2009 period the following data are used:
\begin{itemize}
\item $\widetilde{n}_i$ --the annual number of losses above $\widetilde{u}=$ Euro 10,000 in year $i\in\{1,\ldots,T\}$;
\item $n_i$ -- the annual number of losses above $u=$ Euro 20,000 in year $i\in\{1,\ldots,T\}$;
\item ${S}_i$ -- the sum of losses above the level $u=$ Euro 20,000 in year $i\in\{1,\ldots,T\}$;
\item $M_i$ -- maximum individual loss in year $i\in\{1,\ldots,T\}$.
\end{itemize}

  The hybrid parameter estimation assumes that the frequency and severity distributions are unchanged over 5 years. Then the following statistics are defined:
%\vspace{0.1cm}
%
%%%%%%%%%%%%%%%%%%%%%%%%%%%%%%%%%%%%%%%%%%%%%%%%%%%%%%%%%%%%%%%%%%%%%%%%%%%%%%%
%\noindent \textbf{LDA Model Frequency Based Population Distribution Equations.}\\
%%%%%%%%%%%%%%%%%%%%%%%%%%%%%%%%%%%%%%%%%%%%%%%%%%%%%%%%%%%%%%%%%%%%%%%%%%%%%%%
%In the case of Poisson frequency we have from these definitions a thinned Poisson process with definition for the intensity parameters given by:
\begin{equation}\label{lambda_cond}
\lambda_u:=\mathrm{E}[N|X\ge u]=\lambda (1-F_X(u;\bm\theta)),
\end{equation}
\begin{equation}
\lambda_{\widetilde{u}}:=\mathrm{E}[N|X\ge \widetilde{u}]=\lambda (1-F_X(\widetilde{u};\bm\theta)),
\end{equation}
\begin{equation}\label{mean_cond}
\mu_u:=\mathrm{E}_{\bm\theta}[X|X\ge u],
\end{equation}
which are estimated using observed frequencies $n_i$ and $\widetilde{n}_i$, and aggregated losses $S_i$ as
\begin{equation}\label{populationstat_eq}
\widehat\lambda_u=\frac{1}{T}\sum_{i=1}^T n_i,\quad \widehat\lambda_{\widetilde{u}}=\frac{1}{T}\sum_{i=1}^T \widetilde{n}_i,\quad \widehat\mu_u=\frac{\sum_{i=1}^T S_i}{\sum_{i=1}^T n_i}.
\end{equation}

The conditional mean $\mu_u$ and severity distribution $F_X(\cdot;\bm\theta)$ are known in closed for the selected severity distribution types, see  \cite[p.26, table A.5]{BCBSd2912014}. Then, in the case of Lognormal, Log-gamma, Log-logistic and Pareto-light distributions, the following two equations are solved to find severity parameter estimates $\widehat{\bm\theta}$:
\begin{equation}\label{quantile_cond}
\frac{\widehat\lambda_{\widetilde{u}}}{\widehat\lambda_u}=\frac{1-F_X(\widetilde{u};\widehat{\bm\theta})}{1-F_X(u;\widehat{\bm\theta})}\quad\mbox{and} \quad \widehat\mu_u=\mathrm{E}_{\widehat{\bm\theta}}[X|X\ge u],
\end{equation}
referred to as the \emph{percentile and moment conditions} respectively. Finally, Poisson $\lambda$ parameter  is estimated as
\begin{equation}\label{lambda_estimate_eq}
\widehat{\lambda}=\widehat{\lambda}_u/(1-F_X(u;\widehat{\bm\theta})).
\end{equation}

%
%These population summaries are functions of the unknown parameters of the LDA model to be estimated. Therefore, sample estimates of these population summaries must also be obtained. These were obtained in the OpCar framework for each bank by using only 5 samples, the following quantities are estimated from the data as follows:

%%%%%%%%%%%%%%%%%%%%%%%%%%%%%%%%%%%%%%%%%%%%%%%%%%%%%%%%%%%%%%%%%%%%%%%%%%%%%%
%\noindent \textbf{LDA Model Severity Component Population Distribution Equations.}\\
%%%%%%%%%%%%%%%%%%%%%%%%%%%%%%%%%%%%%%%%%%%%%%%%%%%%%%%%%%%%%%%%%%%%%%%%%%%%%%

In the case of Pareto-heavy severity the percentile condition in (\ref{quantile_cond}) is replaced by the ``maximum heavy condition''
\begin{equation}\label{max_cond1}
F_{X|X>\widetilde{u}}(\widehat\mu_M^{(1)};\widehat{\bm\theta})=\frac{\widetilde{n}}{\widetilde{n}+1},\quad \widehat\mu_M^{(1)}=\max(M_1,\ldots,M_T)
\end{equation}
and in the case of Pareto-medium severity the percentile condition is replaced by the ``maximum medium condition"
\begin{equation}\label{max_cond2}
F_{X|X>\widetilde{u}}(\widehat\mu_M^{(2)};\widehat{\bm\theta})=\frac{\widetilde{n}}{\widetilde{n}+1},\quad \widehat\mu_M^{(2)}=\frac{1}{T}\sum_{j=1}^T M_j.
\end{equation}
Here, $F_{X|X>\widetilde{u}}(\cdot)$ is the distribution of losses conditional to exceed $\widetilde{u}$. Explicit definition of $\widetilde{n}$ is not provided in \cite[annex 2]{BCBSd2912014}, but  it can be reasonable to assume $\widetilde{n}=\frac{1}{T}\sum_{i=1}^T \widetilde{n_i}$ in (\ref{max_cond2}) and $\widetilde{n}=\sum_{i=1}^T \widetilde{n_i}$ in (\ref{max_cond1}).

These maximum conditions are based on the following result and approximation stated in  \cite{BCBSd2912014}. Denote the ordered loss sample $X_{1,n}\le \cdots\le X_{n,n}$, i.e. $X_{n,n}=\max(X_1,\ldots,X_{n})$. Using the fact that $F_X(X_i)=U_i$ is uniformly distributed, we have $\mathrm{E}[F_X(X_{k,n})]=\frac{k}{n+1}$ and thus
\begin{equation}\label{maxcond_eq1}
\mathrm{E}[F_X(X_{n,n})]=\frac{n}{n+1}.
\end{equation}
\noindent Therefore, when $n\rightarrow \infty$ one can expect that
\begin{equation}\label{maxcond_eq2}
\mathrm{E}[F_X(X_{n,n})]\approx F_X[\mathrm{E}(X_{n,n})],
\end{equation}
that gives conditions (\ref{max_cond1}) and (\ref{max_cond2}) when $\mathrm{E}(X_{n,n})$ are replaced by its estimators $\widehat\mu_M^{(1)}$ and $\widehat\mu_M^{(2)}$ and conditional distribution $F_{X|X>\widetilde{u}}(\cdot)$ is used instead of $F_{X}(\cdot)$  to account for loss truncation below $\widetilde{u}$. Here, we would like to note that strictly speaking under the OpRisk settings $n$ is random from Poisson corresponding to the annual number of  events that may have implications to the above maximum conditions. Also note that the distribution of maximum loss  in  the case of Poisson distributed $n$ can be found in closed form, see \cite[section 6.5]{Shevchenko2011}.

%Given these population distribution functions, we can also obtain sampled based estimates as follows. If we denote

We are not aware of results in the literature regarding the properties (such as accuracy, robustness and appropriateness) of estimators $\widehat{\bm\theta}$ calculated in the above described way. Note that it is mentioned in \cite{BCBSd2912014} that if numerical solution for $\widehat{\bm\theta}$ does not exist for a model then this model is ignored.

%We have some doubts for the above estimation using formulas for maximum (\ref{maxcond_eq1},\ref{maxcond_eq2},\ref{maxcond_eq3},\ref{maxcond_eq4}). For a sequence of $n$ iid variables $X_1,\ldots,X_n$ from $F_X(x)$, the distribution of maximum loss $M_n=\max(X_1,\ldots,X_n)$ is $F_{M_n}(x)=(F_X(x))^n$. For a sequence of $X_1,\ldots,X_N$ where $N$ is random from $Poisson(\lambda)$ the distribution of maximum is
%\begin{equation}
%F_{M_N}(x)=\exp\left( -\lambda(1-F_X(x)) \right)
%\end{equation}
%If we want to use average maximum over $T$ years then we need to calculate
%\begin{equation}
%\int x f_{M_N}(x) dx=\int (1-F_{M_N}(x))dx
%\end{equation}
%Also, the distribution of maximum over $T$ years is
%\begin{equation}
%\left(F_{M_N}(x)\right)^T=\exp\left( -\lambda T(1-F_X(x)) \right)
%\end{equation}

%Thus, for estimation of two severity parameters we have 4 conditions (\ref{quantile_cond},\ref{mean_cond},\ref{max_cond1},\ref{max_cond2}). Only two conditions are selected to fit severity model and thus we get 6 possible estimators, all with different properties and different accuracies. Once severity parameters are estimated then frequency $\lambda$ is estimated using (\ref{lambda_cond}).\\

The OpCar framework takes a five-year sample of data to perform sample estimation when fitting the models. We note that making an estimation of this type with only five years of data for the annual loss of the institution and fitting the model to this sample, is going to have a very poor accuracy. This can then easily translate into non-robust results from the OpCar formulation if recalibration of the framework is performed in future.

At this point we emphasize that a five sample estimate is very inaccurate for these quantities, in other words, the estimated model parameters will have a very large uncertainty associated with them. This is particularly problematic for parameters related to kurtosis and tail index in a subexponential severity models. This is probably why the heuristic practically motivated rejection criterion for model fits was applied, to reject inappropriate fits (not in a statistical manner) that did not work due to the very small sample sizes used in this estimation.

To illustrate the extent of the uncertainty present in these five-year sample estimates we provide the following basic case study.
%\vspace{0.1cm}
%\FloatBarrier
%%%%%%%%%%%%%%%%%%%%%%%%%%%%%%%%%%%%%%%%%%%%%%%%%%%%%%%%%%%%%%%%%%%%%%%%%%%%%%
%\begin{example}[Accuracy of the OpCar Calibration Framework]
Consider a Poisson-Lognormal LDA model with parameters $\lambda = 1,000$, $\mu = 10$ and $\sigma = 2$ simulated over $m=1,000$ years. The $VaR_{0.999}$ for this model given by the single loss approximation (\ref{SLALN}) is $4.59\times 10^8$.
Then we calculate population statistics for each year $n_i$, $\widetilde{n}_i$ and ${S}_i$, $i=1,\ldots,m$ and form $T$-year non-overlapping blocks of these statistics from the simulated $m$ years to perform estimation of distribution parameters for each block using percentile and moment conditions (\ref{quantile_cond}). Formally, for each block, we have to solve numerically two equations
\begin{equation}\label{calibration_eqsolve_eq}
\begin{split}
O_1 &:= \Phi\left(\frac{\widehat\mu-\ln(\widetilde{u})}{\widehat\sigma}\right).\left[\Phi\left(\frac{\widehat\mu-\ln(u)}{\widehat\sigma}\right)\right]^{-1} - \frac{\widehat\lambda_{\widetilde{u}}}{\widehat\lambda_{u}} = 0,\\
O_2 &:= \exp\left(\widehat\mu + \frac{\widehat\sigma^2}{2}\right)\Phi\left(\frac{\widehat\mu + \widehat\sigma^2-\ln(u)}{\widehat\sigma}\right)\left[\Phi\left(\frac{\widehat\mu-\ln(u)}{\widehat\sigma}\right)\right]^{-1} - \widehat\mu_u = 0,
\end{split}
\end{equation}
to find severity parameter estimates $\widehat{\mu}$ and $\widehat{\sigma}$ which are then substituted into (\ref{lambda_estimate_eq}) to get estimate $\widehat\lambda$. Here, $\widehat\lambda_{\widetilde{u}}$, ${\widehat\lambda_{u}}$ and $\widehat\mu_u$ are observed statistics (\ref{populationstat_eq}) for a block.

%according to:
%\begin{equation}
%\lambda = \frac{\widehat\lambda_u}{1 - F_X\left(u;\widehat{\mu},\widehat{\sigma}\right)}.
%\end{equation}

 This system of non-linear equations may not have a unique solution or a solution may not exist. Thus, to find approximate solution, we consider two different objective  functions:
 \begin{itemize}
\item objective  function 1: a univariate objective function given by $O_1^2 + O_2^2$ that we minimize to find the solution; and
\item objective  function 2: a multi-objective function to obtain the Pareto optimal solution by finding the solution such that $|O_1|$ and $|O_2|$ cannot be jointly better off.
\end{itemize}
 In both cases, a simple grid search over equally spaced values of $\widehat\mu$ and $\widehat\sigma$ was used to avoid any other complications that may arise with other optimization techniques -- this will lead to the most robust solution which is not sensitive to gradients or starting points. A summary of the results for parameter estimates and corresponding $VaR_{0.999}$ in the case of 5-year blocks (i.e. $[m/T]=200$ independent blocks) is provided in Table \ref{TableEmpEstPerforance}.

 \begin{table}[!h]
\centering
\captionsetup{width=0.95\textwidth}
\caption{\footnotesize{Mean of OpCar model parameter estimates over 200 independent 5-year blocks if data are simulated from $Poisson(1,000)-Lognormal(10,2)$. The mean of sample statistics $\widehat{\lambda}_u$, $\widehat{\lambda}_{\tilde{u}}$ and $\widehat{\mu}_u$ over the blocks used for parameter estimates are: 654(11), 519(10) and $3.03(0.27)\times 10^5$ respectively with corresponding standard deviations provided in brackets next to the mean. The square root of mean squared error of the parameter estimator is provided in brackets next to the mean.}}
\label{TableEmpEstPerforance}
{\footnotesize{
\begin{tabular}{ccc}
\toprule
\textbf{parameter estimate} & \textbf{objective function 1}    & \textbf{objective function 2} \\ \midrule
\multicolumn{3}{c}{Grid 1: $\widehat\mu\in[8,12]$, $\widehat\sigma\in [1,3]$, $\delta\widehat\mu=0.05$, $\delta\widehat\sigma=0.05$} \\ \midrule
$\widehat\mu$ 						        & $10.12(1.26)$       		 &  $9.35(1.03)$ \\
$\widehat\sigma$ 					        & $1.88(0.46)$ 						 &  $2.16(0.28)$ \\
$\widehat\lambda$				  & $3248(5048)$   					 &  $959(290)$ \\
$\widehat{VaR}_{0.999}$   & $10.6(13.4)\times 10^8$ &  $4.70(0.88)\times 10^8$\\ \midrule
\multicolumn{3}{c}{Grid 2: $\widehat\mu\in[6,14]$, $\widehat\sigma\in [0.5,3.5]$, $\delta\widehat\mu=0.05$, $\delta\widehat\sigma=0.05$} \\ \midrule
$\widehat\mu$ 						        & $9.79(1.94)$       		 &  $8.85(1.82)$ \\
$\widehat\sigma$ 					        & $1.88(0.71)$ 						 &  $2.26(0.48)$ \\
$\widehat\lambda$				  & $1.18(11.1)\times 10^8 $  					 &  $1.5(21)\times 10^7$ \\
$\widehat{VaR}_{0.999}$   & $3.84(36.8)\times 10^{13}$ &  $4.34(61)\times 10^{12}$\\
\bottomrule
\end{tabular}
}}
\end{table}

 The results are presented for two search grids with different bounds but the same spacing between the grid points $\delta\widehat\mu=0.05$ and $\delta\widehat\sigma=0.05$.
 Here, $VaR_{0.999}$ is calculated using single loss approximation (\ref{SLALN}). \emph{\textbf{These results clearly show that this estimation procedure is not appropriate}}.
We note that in some instances (for some blocks), both objective functions produce the same parameter estimates for common sample estimate inputs $\widehat\lambda_{\widetilde{u}}$, $\widehat\lambda_{u}$ and $ \widehat\mu_u$, and these estimates are close to the true values, though this was not systematically the case. We suspect that the reason for this is that in some instances there may be no solution existing or multiple solutions and this then manifest in different solutions for the two objective functions or inappropriate solutions.

 This is a serious concern for the accuracy of the findings from this approach -- it suggests and is primarily a key reason why other approaches to calibration at more granular levels (where more data is available) are often used. Or in cases where events are very rare, why alternative sources of data such as KRI, KPI, KCI and expert opinions should be incorporated into the calibration of the LDA model.

%\FloatBarrier

%%%%%%%%%%%%%%%%%%%%%%%%%%%%%%%%%%%%%%%%%%%%%%%%%%%%%%%%%%%%%%%%%%%%%%%%%%%%%%
\subsection{Capital Estimation}\label{CapitalEstimationOpCar_sec}
%%%%%%%%%%%%%%%%%%%%%%%%%%%%%%%%%%%%%%%%%%%%%%%%%%%%%%%%%%%%%%%%%%%%%%%%%%%%%%
The second stage of the OpCar framework is to take the fitted model parameters for the LDA model severity and frequency models at group or institution level and then calculate the capital.
The approach to capital calculation under the OpCar analysis involves the so-called single loss approximation (SLA). Here we note that it would have been more accurate and not difficult to perform the capital calculations numerically using methods such as Panjer recursion, fast Fourier transform (FFT) or Monte Carlo; see detailed discussion on such approaches provided in \cite{CruzPetersShevchenko2015}.

Instead, the BCBS decided upon the following SLA to estimate the 0.999 quantile of the annual loss
\begin{equation}\label{SLA_eq1}
F_Z^{-1}(\alpha)\approx F_X^{-1}\left(1-\frac{1-\alpha}{\lambda}\right)+(\lambda-1)\mathrm{E}[X],
\end{equation}
which is valid asymptotically for $\alpha\rightarrow 1$ in the case of subexponential severity distributions. Here, $F_X^{-1}(\cdot)$ is the inverse of the distribution function of random variable $X$. It is important to point that
correct SLA (in the case of finite mean subexponential severity and Poisson frequency) is actually given by slightly different formula
\begin{equation}\label{SLA_eq2}
F_Z^{-1}(\alpha) = F_X^{-1}\left(1-\frac{1-\alpha}{\lambda}\right)+ \lambda \mathrm{E}[X] +o(1).
\end{equation}
This is a reasonable approximation but it is important to note that its accuracy depends on distribution type and values of distribution parameters and further higher order approximations are available; see detailed discussion in \cite[section 8.5.1]{PetersShevchenko2015} and the tutorial paper \cite{peters2013understanding}.

%\begin{remark}
%We also point out that numerous asymptotic approximations are available for quantiles and risk measures of an LDA annual loss distribution. Each approximation comes with different assumptions about dependence and about the tail behaviour of the severity model, see discussions in detail in \cite{peters2013understanding} and \cite{PetersShevchenko2015}. Since a particular version of such approximations is assumed in the OpCar framework, and critically it is used directly for calibration of the model and all resulting analysis is based on a critical approximation made at this point, we first detail what exactly is the approximation made. We then explain when such an approximation is accurate and when it is inaccurate for a range of model parameters and models that were used in the OpCar framework analysis. This will shed light we believe on possible problems with the resulting SMA framework that evolved out of the OpCar set-up.
%\end{remark}

     To illustrate the accuracy of this first order approximation for the annual loss VaR at a level of 99.9\%, consider the $Poisson(\lambda)-Lognormal(\mu,\sigma)$ model with parameters $\lambda =\{10,100,1,000\}$, $\mu = 3$ and $\sigma = \left\{1,2\right\}$, note that parameter $\mu$ is a scale parameter and will not have impact on relative differences in VaR. Then we calculate the VaR of the annual loss from an LDA model for each of the possible sets of parameters  using Monte Carlo simulation of $10^7$ years that gives very good accuracy. We then evaluate the SLA approximation for each set of parameters and a summary of the findings is provided in Table \ref{TableOpCarAcc}.

\begin{table}[!h]
\centering
\captionsetup{width=0.95\textwidth}
\caption{\footnotesize{Accuracy of the SLA approximation of $VaR_{0.999}$ in the case of $Poisson(\lambda)-Lognormal(\mu,\sigma)$ model with the scale parameter $\mu=3$. $\Delta$VaR is the difference between the SLA approximation and Monte Carlo estimate (standard error of the Monte Carlo approximation is in brackets next to the estimate).  $\epsilon(\%)$ is relative difference between the SLA and Monte Carlo estimates.}}
\label{TableOpCarAcc}
{\footnotesize{
\begin{tabular}{lllll}
\toprule
\textbf{Parameters} & MC VaR & SLA VaR & $\Delta$VaR & $\epsilon(\%)$ \\ \midrule
$\lambda = 1,000,\, \sigma = 1 $& $3.88\times 10^4 (0.02\%)$ & $3.54\times 10^4$  & $-3.4\times10^3$  & $-8.7\%$ \\
$\lambda = 1,000,\,  \sigma = 2 $  & $4.24\times 10^5 (0.28\%)$ & $4.19\times 10^5$ & $-5.3\times10^3$ & $-1.3\%$ \\
$\lambda = 100,\,  \sigma = 1 $  & $5.42\times 10^3(0.06\%)$ & $4.74\times 10^3$ & $-6.8\times 10^2$  &
$-12.5\%$ \\
$\lambda = 100,\,  \sigma = 2 $  & $1.17\times 10^5 (0.4\%)$ & $1.17\times 10^5$ & $-6.2\times10^2$ & $-0.52\%$ \\
$\lambda = 10,\,  \sigma = 1 $ & $1.27\times 10^3 (0.15\%)$ & $1.16\times 10^3$ &$-1.1\times10^2$  & $-8.7\%$ \\
$\lambda = 10,\,  \sigma = 2 $ & $3.57\times 10^4(0.52\%)$& $3.56\times 10^4$ & $-0.8\times10^2$ & $-0.23\%$ \\
\bottomrule
\end{tabular}
}}
\end{table}

The results indicate that though the SLA accuracy is good, the error can be significant for the parameters of interest for OpCar modelling. This can lead to material impact on the accuracy of the parameter estimation in the subsequent regression undertaken by the OpCar approach and the resulting SMA formula.

%%%%%%%%%%%%%%%%%%%%%%%%%%%%%%%%%%%%%%%%%%%%%%%%%%%%%%%%%%%%%%%%%%%%%%%%%%%%%%
\subsection{Model Selection and Model Averaging}
%%%%%%%%%%%%%%%%%%%%%%%%%%%%%%%%%%%%%%%%%%%%%%%%%%%%%%%%%%%%%%%%%%%%%%%%%%%%%%
The OpCar methodology attempts to fit six different severity models to the data as described in the previous section that generates up to six 0.999 VaR SLA estimates for each bank depending on the models survived the imposed ``filters". Then the final estimate of the 0.999 VaR is found as an average of VaR estimates from the survived models.

We would like to comment on the ``filters'' that were used to select the models to be used for a bank, after the estimation procedures were completed. These heuristic adhoc filters are not based on statistical theory and comprised of the following rules:
\begin{itemize}
\item whether the proportion of losses above Euro 20,000 was within a certain range (1\%--40\%);
\item whether the ratio between loss frequency and total assets was within a certain range (0.1--70 losses per billion Euro of assets);
\item whether the model estimation (outlined in Section \ref{ParamEstimationOpCar_sec}) based on iterative solution converged.
\end{itemize}
We believe that in addition to practical considerations, other more rigorous statistical approaches to model selection could also be considered. For instance, \cite{dutta2006tale} discuss the importance of fitting distributions that are flexible but appropriate for the accurate modelling of OpRisk data; they focussed on the following five simple attributes in deciding on a suitable statistical model for the severity distribution.
\begin{enumerate}
\item \emph{Good Fit}. Statistically, how well does the model fit the data?
\item \emph{Realistic}. If a model fits well in a statistical sense, does it generate a loss distribution with
a realistic capital estimate?
\item \emph{Well Specified}. Are the characteristics of the fitted data similar to the loss data and logically
consistent?
\item \emph{Flexible}.How well is the model able to reasonably accommodate a wide variety of empirical
loss data?
\item \emph{Simple}. Is the model easy to apply in practice, and is it easy to generate random numbers
for the purposes of loss simulation?
\end{enumerate}
Furthermore, in \cite[chapter 8]{CruzPetersShevchenko2015} there is a detailed description of appropriate model selection approaches that can be adopted in OpRisk modelling, specifically developed for this setting that involve rigorous statistical methods, that avoid heuristic methods.

The results of the heuristic filters applied in OpCar excluded a large number of distribution fits from the models for each institution. It was stated in \cite[page 22]{BCBSd2912014} that \emph{``OpCar calculator was run and validated on a sample of 121 out of 270 QIS banks which were able to provide data on operational risk losses of adequate quality"} and ``\emph{four of the distributions (log normal, log gamma, Pareto medium, and Pareto heavy) were selected for the final OpCar calculation around 20\% of the time or less}''.

%This result on its own is clearly at odds with OpRisk best practice since it was clear from the survey of AMA validated banks, performed by BIS, which showed that Lognormal and Pareto were by far the most popularly and regularly used models in OpRisk practice. Furthermore, these models are from the family of sub-exponential loss models - which are typically considered as highly desirable.

 %in the regulation see ....  the Committee's ``Operational risk – supervisory guidelines for the advanced measurement approaches (July 2011)", ``in such cases the use of so-called sub-exponential
%distributions is highly recommended" (emphasis added; page 40).

%%%%%%%%%%%%%%%%%%%%%%%%%%%%%%%%%%%%%%%%%%%%%%%%%%%%%%%%%%%%%%%%%%%%%%%%%%%%%%
\subsection{OpCar Regression Analysis}
%%%%%%%%%%%%%%%%%%%%%%%%%%%%%%%%%%%%%%%%%%%%%%%%%%%%%%%%%%%%%%%%%%%%%%%%%%%%%%
    Finally we discuss aspects of regression in OpCar methodology based on OpCar parameter and VaR estimation outlined in the previous section. Basically for each bank's approximated  capital, the regression is performed against a range of factors in a linear and non-linear regression formulation. Given the potential for significant uncertainty in all the aspects of the OpCar framework presented above, we argue that results from this remaining analysis may be spurious or biased and we would recommend further study on this aspect.
    %Since, one would expect the accuracy of the parameters obtained by the OpCar calibration procedure to be affected in an amplified manner by the approximation errors introduced as shown in the above study.
    Furthermore, these approximation errors will propagate into the regression parameter estimation in a non-linear manner making it difficult to directly determine the effect of such inaccuracies.

The regression models considered came in two forms, linear and non-linear regressions. Given the sample of $J$ banks on which to perform the regression, consider $\left(Y_j,X_{1,j},\ldots,X_{20,j}\right)$, $j=1,\ldots,J$, where
$Y_j$ is the $j$-th banks capital (dependent variable) obtained from the two stage procedure described in Sections \ref{CapitalEstimationOpCar_sec} and \ref{ParamEstimationOpCar_sec}. Here, $X_{i,j}$ is the $i$-th factor or covariate (independent variable) derived from the balance sheet and income statement of the $j$-th bank. OpCar methodology considered 20 potential covariates.
\emph{Only one observation of dependent variable $Y_j$ per bank was used and thus only one observation for each independent variable $X_{i,j}$ was needed and approximated by the average over QIS reporting years.}

Then the linear regression model considered in OpCar was specified by the model
\begin{equation} \label{generalizedLinearRegressions}
Y_j = b_0 + \sum_{i=1}^{20} b_i X_{i,j} + \epsilon_j,
\end{equation}
with i.i.d. errors $\epsilon_1,\ldots,\epsilon_J$ are from Normal distribution with zero mean and the same variance.

In this particular practical application  it is unlikely that the regression assumption of homoskedastic error variance is appropriate. We comment that the failure of this assumption, which is likely to be true given the heterogeneity of the banks considered, would have caused spurious results in the regression analysis. The comment on this regard is that ``outliers'' were removed in the analysis, no indication of how or what were considered outliers was mentioned. This again would have biased the results. We would have suggested to not arbitrarily try to force homoskedastic errors but to instead consider weighted least squares if there was a concern with outliers.

The regression analysis undertaken by the OpCar analysis and then utilised as a precursor to the SMA formulation implicitly assumes that all bank capital figures, used as responses (dependent variables) in the regression analysis against the factors (independent variables), such as BI for each bank, come from a common population. This is seen by the fact that the regression is done across all banks jointly, thereby assuming a common regression error distribution type. Clearly, this would probably not be the case in practice, we believe that other factors such as banking volume, banking jurisdiction, banking practice and banking risk management governance structures etc. can have significant influences on this potential relationship. To some extend the BI is supposed to capture aspects of this but it cannot capture all of these aspects. Therefore, it maybe the case that the regression assumptions about the error distribution, typically being i.i.d. zero mean homoskedastic (constant) variance and Normally distribution would probably be questionable. Unfortunately, this would then directly affect all the analysis of significance of the regression relationships, the choice of the covariates to use in the model, etc.

It is very odd to think that in the actual OpCar analysis,
the linear regression model (\ref{generalizedLinearRegressions}) was further reduced to the simplest type of linear regression model in the form of a simple linear model sub-family given by
\begin{equation}
Y_j = b_0 + b_i X_{i,j} + \epsilon_j,
\end{equation}
i.e. only simple linear models and not generalized linear model types  were considered. This is unusual as the presence of multiple factors and their interactions often can significantly improve the analysis and it is simple to perform estimation in such cases. It is surprising to see this very limiting restriction in the analysis.

The second form of regression model considered in OpCar was non-linear and given by the functional relationships
\begin{equation}
\begin{split}
R(x) &= xF(x),\\
\frac{d}{dx}R(x) &= F(x) + xF'(x),\\
\frac{d^2}{dx^2}R(x) &= 2F'(x) + xF''(x),\\
\end{split}
\end{equation}
where $x$ is the proxy indicator, $R(x)$ represents the total OpRisk requirement (capital) and $F(x)$ is the functional coefficient relationship for any level $x$. It is assumed that $F(\cdot)$ is twice differentiable. The choice of function $F(x)$ selected was given by
\begin{equation}
F(x) = \theta \frac{(x-A)^{1-\alpha}}{1-\alpha},
\end{equation}
with $\alpha \in [0,1]$, $\theta \geq 0$ and $A \leq 0$.

This model and the way it is described in the Basel consultative document is incomplete in the sense that it failed to adequately  explain how multiple covariates were incorporated into the regression structure. It seems that again only a single covariate is considered one at a time -- we again emphasize that this is a very limited and simplistic approach to performing such analysis. There are standard software R packages that would have extended this analysis to multiple covariate regression structures which we argue would have been much more appropriate.

Now, in the non-linear regression model if one takes $R(x_i) = Y_i$, i.e. the $i$-th banks capital figure, this model could in principle be reinterpreted as a form of quantile regression model such as those discussed recently in \cite{CruzPetersShevchenko2015}. In this case the response is the quantile function and the function $F$ would have been selected as a transform of a quantile error function, such as the class of Tukey transforms discussed in \cite{peters2016estimating}.

The choice of function $F(x)$ adopted by the modellers was just a translated and scaled Power quantile error function of the type discussed in \cite[equation 14]{dong2015risk}.
When interpreting the non-linear model in the form of a quantile regression it is documented in several places (see discussion in \cite{peters2016estimating}) that least squares estimation is not a very good choice for parameter estimation for such models. Yet, this is the approach adopted in the OpCar framework.

Typically, when fitting quantile regression models one would instead use a loss function corresponding to minimizing the expected loss of $Y-u$ with respect to $u$ according to
\begin{equation}
\text{min}_u\mathrm{E}[\rho_{\tau}(Y-u)],
\end{equation}
where $\rho_{\tau}(y)$ is given by
\begin{equation}
\rho_{\tau}(y) = y(\tau - 1_{\{y < 0\}}).
\end{equation}

Since the OpCar framework is only defined at the institutional level, this means that effectively the modelling of the regression framework cannot easily incorporate in a natural way BEICFs such as KRI, KCI and KPI since these measures are typically recorded at a more granular level than the institution level. Instead, new OpRisk proxies and indicators were created under OpCar methods based on balance sheet outputs.

In fact, 20 proxy indicators were developed from the BCBS 2010 QIS data balance sheets and income statements of the participating banks selected. These included refinements related to gross income, total assets, provisions, administrative costs as well as alternative types of factors.
Unfortunately the exact choice of 20 indicators used were not explicitly stated in detail in the OpCar document \citep[annex 3]{BCBSd2912014}, making it difficult to discuss the pros and cons of the choices made. Nor was there a careful analysis undertaken of whether such factors selected could have produced possible collinearity issues in the regression design matrix. For instance, if you combine one covariate or factor with another factor derived from this one or strongly related to it (as seemed to be suggested in the cases with the GI based factors), then the joint regression modelling with both factors will lead to an increased variance in parameter estimation and misguided conclusions about significance (statistical and practical) with regard to the factors selected in the regression model.
%%%%%%%%%%%%%%%%%%%%%%%%%%%%%%%%%%%%%%%%%%%%%%%%%%%%%%%%%%%%%%%%%%%%
\section{Proposition: a Standardization of AMA}\label{StandardisationAMA}
%%%%%%%%%%%%%%%%%%%%%%%%%%%%%%%%%%%%%%%%%%%%%%%%%%%%%%%%%%%%%%%%%%%%
%%%%%%%%%%%%%%%%%%%%%%%%%%%%%%%%%%%%%%%%%%%%%%%%%%%%%%%%%%%%%%%%%%%%
%In the following section the recommendations made are based on detailed discussions proposed in \cite{CruzPetersShevchenko2015}, \cite{PetersShevchenko2015} and the preprint \cite{PetersEtAl2016}.

\noindent SMA cannot be considered as an alternative to AMA models.  We suggest that AMA should not be discarded, but instead could be improved by addressing its current weaknesses. \emph{\textbf{It should be standardized!}} Details of how a rigorous and statistically robust standardization can start to be considered, with practical considerations, are suggested below.

Rather than discarding all OpRisk modelling as allowed under the AMA, instead the regulator could make a proposal to standardize the approaches to modelling based on the accumulated knowledge to date of OpRisk modelling practice.
We propose one class of models that can act in this manner and allow one to incorporate the key features offered by AMA LDA type models which involve internal data, external data, BEICF's and scenarios, with other important information on factors that the SMA method and OpCar approaches have tried to achieve but failed. As has already been noted, one issue with the SMA and OpCar approaches is that they try to model all OpRisk processes at the institution or group level with a single LDA model and simplistic regression structure, this is bound to be problematic due to the very nature and heterogeneity of OpRisk loss processes. In addition it fails to allow for incorporation of many important OpRisk loss process explanatory information sources such as BEICFs which are often no longer informative or appropriate to incorporate at institution level, compared to individual business line/event type (BL/ET) level.  				

 A standardization of the AMA internal models will remove the wide range of heterogeneity in model type. Here, our recommendation involves a bottom up modelling approach where for each BL/ET  OpRisk loss process we model the severity and frequency components in an LDA structure. It can be comprised of a hybrid LDA model with factor regression components, allowing to include the factors driving OpRisks in the financial industry, at a sufficient level of granularity, while utilizing a class of models known as the \emph{Generalized Additive Models for Location, Shape and Scale} (GAMLSS) in the severity and frequency aspects of the LDA framework. The class of GAMLSS models can be specified to make sure that the severity and frequency families are comparable across institutions, allowing both risk-sensitivity and capital comparability. We recommend in this regard Poisson and Generalized Gamma classes for the family of frequency and severity model as these capture all typical ranges of loss model used in practice over the last 15 years in OpRisk, including Gamma, Weibull, Lognormal and Pareto type severities.

\vspace{0.2cm}

\noindent{\textbf{Standardizing Recommendation 1.}}
This leads us to the first standardizing recommendation relating to the level of granularity of modelling in OpRisk. The level of granularity of the modelling procedure is important to consider when incorporating different sources of OpRisk data such as BEICF's and scenarios and this debate has been going for the last 10 years, with discussion on bottom-up versus top-down based OpRisk modelling, see overview in \cite{CruzPetersShevchenko2015} and \cite{PetersShevchenko2015}. We advocate that a bottom-up based approach be recommended as the standard modelling structure as it will allow for greater understanding and more appropriate model development of the actual loss processes under study. Therefore, we argue that sticking with the 56 BL/ET structure of Basel II is in our opinion best for a standardizing framework with a standard aggregation procedure to institution level/group level. We argue that alternatives such as the SMA and OpCar approaches that are trying to model multiple different featured loss processes combined into one loss process at the institution level will be bound to fail as they need to capture high frequency events, as well as high severity events. This in principle is very difficult if not impossible to capture with a single LDA model at institution level and should be avoided. Furthermore, such a bottom-up approach allows for greater model interpretation and incorporation of OpRisk loss data such as BEICFs.

\vspace{0.2cm}

\noindent{\textbf{Standardizing Recommendation 2.}}
This brings us to our second recommendation for standardization in OpRisk modelling. Namely, we propose to standardize the modelling class to remove the wide range of heterogeneity in model type. We propose a standardization that involves a bottom up modelling approach where each BL/ET level of OpRisk loss process we model the severity and frequency components in an LDA structure which is comprised of a hybrid LDA model with factor regression components. The way to achieve this is to utilise a class of GAMLSS regression models for the severity and frequency model calibrations.
That is two GAMLSS regression models are developed, one for the severity fitting and the other for the frequency fitting. This family of models is flexible enough in our opinion to capture any type of frequency or severity model that may be observed in practice in OpRisk data whilst incorporating factors such as BEICFs (Key Risk Indicators, Key Performance Indicators, Key Control Indicators) naturally into the regression structure. This produces a class of hybrid factor regression models in an OpRisk LDA family of models that can easily be fit, simulated from and utilised in OpRisk modelling to aggregate to the institution level. Furthermore, as more data years of history become available, the incorporation of time-series structure in the severity and frequency aspects of each loss process modelling can be naturally incorporated in a GAMLSS regression LDA framework.

\vspace{0.2cm}

\noindent{\textbf{Standardizing Recommendation 3.}}	
The class of models considered for the conditional response in the GAMLSS severity model can be standardized. There are several possible examples of such models that may be appropriate \cite{chavez2015extreme} and \cite{ganegoda2013scaling}. However, we advocate for the severity models that the class of models be restricted in regulation to one family, the Generalized Gamma family of models, %see details in \cite{PetersEtAl2016}
where these models are developed in an LDA hybrid factor GAMLSS model. Such models are appropriate for OpRisk as they admit special members which correspond to the Lognormal, Pareto, Weibull and Gamma. All of these models are popular OpRisk severity models used in practice and represent the range of best practice by AMA banks as observed in the recent survey \cite{BCBS160b}. Since the Generalized Gamma family contains all these models as special sub-cases it means that banks would only have to ever fit one class of severity model to each BL/ET LDA severity profile. Then, the most appropriate family member would be resolved in the fitting through the estimation of the shape and scale parameters, in such a manner that if a Lognormal model was appropriate it would be selected, where as if a Gamma model were more appropriate it would also be selected from one single fitting procedure.
Furthermore, the frequency model could be standardized as a Poisson GAMLSS regression structure as the addition of explanatory covariates and time varying and possible stochastic intensity allow for a flexible enough frequency model for all types of OpRisk loss process.

\vspace{0.2cm}

\noindent{\textbf{Standardizing Recommendation 4.}}
The fitting of these models should be performed in a regression based manner in the GAMLSS framework, which incorporates truncation and censoring in a penalized maximum likelihood framework, see \cite{StasinopoulosRigby2007}. We believe by standardizing the fitting procedure to one that is statistically rigorous, well understood in terms of the estimator properties and robust when incorporating a censored likelihood appropriately, will remove the range of heuristic practices that has arisen in fitting models in OpRisk. The penalized regression framework, based on L1 parameter penalty will also allow for shrinkage methods to be used to select most appropriate explanatory variables in the GAMLSS severity and frequency regression structures.

\vspace{0.2cm}

\noindent{\textbf{Standardizing Recommendation 5.}}
The standardization in form of Bayesian versus Frequentist type models be left to the discretion of the bank to decide which version is best for their practice. However, we note that under a Bayesian formulation, one can adequately incorporate multiple sources of information including expert opinion and scenario based data, see discussions in \cite{CruzPetersShevchenko2015} and \cite{PetersShevchenkoWuthrich2009} and \cite{ShevchenkoWuthrich2006}.

\vspace{0.2cm}

\noindent{\textbf{Standardizing Recommendation 6.}}
The sets of BEICFs and factors to be incorporated into each BL/ET LDA factor regression model for severity and frequency should be specified by the regulator. There should be a core set of factors to be incorporated by all banks which include BEICFs and other factors to be selected. The following types of KRI's categories can be considered in developing the core family of factors (see \cite{Chapelle2013}).
\begin{itemize}
\item \emph{Exposure Indicators}: any significant change in the nature of the business environment and in its exposure to critical stakeholders or critical resources. Flag any change in the risk exposure.
    \item \emph{Stress Indicators}: any significant rise in the use of resources by the business, whether human or material. Flag any risk rising from overloaded humans or machines.
\item \emph{Causal Indicators}: metrics capturing the drivers of key risks to the business. The core of preventive KRIs.
\item \emph{Failure Indicators}: poor performance and failing controls are strong risk drivers. Failed KPIs and KCIs.
\end{itemize}

In this approach, a key difference is that instead of fixing the regression coefficients for all banks (as is the case for SMA and OpCar) pretending that all banks have the same regression relationship as the entire banking population,  one should standardize the class of factors. Specify explicitly how they should be collected, the frequency and then specify that they should be incorporated in the GAMLSS regression. This will allow each bank to then calibrate the regression model to their loss experience through a rigorous penalized Maximum Likelihood procedure. With strict criterion on cross validation based testing on the amount of penalization admitted in the regression when shrinking factors out of the model. This approach has the advantage that banks will not only start to better incorporate in a structured and statistically rigorous manner the BEICF information into OpRisk models, but they will be forced to better collect and consider such factors in a principled manner.

%%%%%%%%%%%%%%%%%%%%%%%%%%%%%%%%%%%%%%%%%%%%%%%%%%%%%%%%%%%%%%%%%%%%
%%%%%%%%%%%%%%%%%%%%%%%%%%%%%%%%%%%%%%%%%%%%%%%%%%%%%%%%%%%%%%%%%%%%
\section{Conclusions}
%%%%%%%%%%%%%%%%%%%%%%%%%%%%%%%%%%%%%%%%%%%%%%%%%%%%%%%%%%%%%%%%%%%%
%%%%%%%%%%%%%%%%%%%%%%%%%%%%%%%%%%%%%%%%%%%%%%%%%%%%%%%%%%%%%%%%%%%%
\noindent In this paper we discussed and studied the weaknesses of the SMA formula for OpRisk capital recently proposed by the Basel Committee to replace AMA and other current approaches. We also outlined the issues with the closely related OpCar model which is the precursor of the SMA. There are significant potential problems with the use of SMA such as capital instability, risk insensitivity and capital super-additivity and serious concerns regarding estimation of this model. We advocate standardisation of AMA rather than its complete removal and provide several recommendations  based on our experience with OpRisk modelling.

\section{Acknowledgment}
Pavel Shevchenko acknowledges the support of the
Australian Research Council's Discovery Projects funding scheme (project number: DP160103489). We also would like to thank many OpRisk industry practitioners and academics  in Australia, Europe, UK and USA, and recent OpRisk Europe 2016 conference in London for useful discussions, and for sharing the views and findings  that has helped in the development and presentation of ideas within this paper.

\newpage
%\bibliographystyle{model2-names}
%\biboptions{authoryear}
%\bibliography{bibliography}
{\footnotesize{
\bibliographystyle{chicago}
\bibliography{bibliography}
}}

%\newpage

%%%%%%%%%%%%%%%%%%%%%%%%%%%%%%%%%%%%%%%%%%%%%%%%%%%%%%%%%%%%%%%%%%%%
%%%%%%%%%%%%%%%%%%%%%%%%%%%%%%%%%%%%%%%%%%%%%%%%%%%%%%%%%%%%%%%%%%%%
%\appendix
%%%%%%%%%%%%%%%%%%%%%%%%%%%%%%%%%%%%%%%%%%%%%%%%%%%%%%%%%%%%%%%%%%%%
%%%%%%%%%%%%%%%%%%%%%%%%%%%%%%%%%%%%%%%%%%%%%%%%%%%%%%%%%%%%%%%%%%%%

%%%%%%%%%%%%%%%%%%%%%%%%%%%%%%%%%%%%%%%%%%%%%%%%%%%%%%%%%%%%%%%%%%%%
%%%%%%%%%%%%%%%%%%%%%%%%%%%%%%%%%%%%%%%%%%%%%%%%%%%%%%%%%%%%%%%%%%%%

\end{document}